\documentclass[12pt, draftclsnofoot, onecolumn]{IEEEtran}
\IEEEoverridecommandlockouts
\usepackage{amsmath,graphicx,amssymb,mathtools,bm}
\usepackage{subfigure}
\usepackage{hyperref}
\usepackage{cite}
\usepackage{amsmath,amssymb,amsfonts}
\usepackage{algorithmic}
\usepackage{textcomp}
\usepackage{xcolor}
\usepackage{verbatim}  
\usepackage{bm}  
\usepackage{mathrsfs} 
\usepackage{algorithm} 
\usepackage{algorithmic} 
\usepackage{booktabs}
\usepackage{textcomp}  
\usepackage{multirow}  
\usepackage{lettrine}   
\usepackage{color}  
\usepackage{amsmath}
\usepackage{amssymb}
\usepackage{amsmath}
\UseRawInputEncoding
\usepackage[justification=centering]{caption}

\def\BibTeX{{\rm B\kern-.05em{\sc i\kern-.025em b}\kern-.08em
    T\kern-.1667em\lower.7ex\hbox{E}\kern-.125emX}}
\begin{document}

\title{  Reconfigurable Intelligent Surface-Aided Full-Duplex mmWave MIMO: Channel Estimation, Passive and Hybrid Beamforming
}

\author{Songjie Yang, Wanting Lyu, Yunis Xanthos,  Zhongpei Zhang,~\IEEEmembership{Member,~IEEE}, \\ Chadi Assi,~\IEEEmembership{Fellow,~IEEE}, and Chau Yuen,~\IEEEmembership{Fellow,~IEEE}


\thanks{S. Yang, W. Lyu, Y. Xanthos, and Z. Zhang are with the National Key Laboratory of Science and Technology on Communications, University of Electronic Science and Technology of China (UESTC), Chengdu 611731, China. (e-mail:
	yangsongjie@std.uestc.edu.cn;	lyuwanting@yeah.net; xiuyue12345678@163.com;
	zhangzp@uestc.edu.cn). C. Assi is with Concordia University, Montreal, Quebec, H3G 1M8, Canada (email:assi@ciise.concordia.ca). C. Yuen is with the Engineering Product Development (EPD) Pillar, Singapore University of Technology and Design,
	Singapore 487372 (e-mail: yuenchau@sutd.edu.sg).

}
}
\maketitle
\begin{abstract}
 Millimeter wave (mmWave) full-duplex (FD) is a promising technique for improving capacity by maximizing the utilization of both time and the rich mmWave frequency resources. Still, it has restrictions due to FD self-interference (SI) and mmWave's limited coverage.
  Therefore, this study dives into FD mmWave MIMO with the assistance of reconfigurable intelligent surfaces (RIS) for capacity improvement. First, we demonstrate the angular-domain reciprocity of FD antenna arrays under the far-field planar wavefront assumption. Accordingly, a strategy for joint downlink-uplink (DL-UL) channel estimation is presented. For estimating the SI channel, the direct channel, and the cascaded channel, the Khatri-Rao product-based compressive sensing (KR-CS), distributed CS (D-CS), and two-stage multiple measurement vector-based D-CS (M-D-CS) frameworks are proposed, respectively. Additionally, we propose a passive beamforming optimization solution based on the angular-domain cascaded channel. With hybrid beamforming architectures, a novel hybrid weighted minimum mean squared error method for SI cancellation (H-WMMSE-SIC) is proposed. Simulations have revealed that joint DL-UL processing significantly improves estimation performance in comparison to separate DL/UL channel estimation. Particularly, when the interference-to-noise ratio is less than 35 dB, our proposed H-WMMSE-SIC offers spectral efficiency  performance comparable to fully-digital WMMSE-SIC. Finally, the computational complexity is analyzed for our proposed methods.

\end{abstract}
\vspace{-0.5cm}
\begin{IEEEkeywords}
Full-duplex, millimeter wave, reconfigurable intelligent surface, channel estimation, beamforming.  
\end{IEEEkeywords}
\vspace{-0.5cm}
\section{Introduction}
To meet the exponential expansion of wireless data demand, high frequency technologies with abundant bandwidth have become indispensable. Besides, the full-duplex (FD) technique, capable of doubling the spectral efficiency (SE) and lowering latency by simultaneously transmitting and receiving data on the same frequency band, has been widely investigated over these years. With rich frequency resources of millimeter wave (mmWave) bands,  
 FD has the potential to bring a significant SE improvement for mmWave communications to support various high-rate services. 
Recently, FD has brought new opportunities for a variety of applications, e.g., wireless power transfer \cite{REC1}, device-to-device systems \cite{STAR}, and integrated access and backhaul \cite{BF6}. 
 Despite the fact that FD offers a substantial advantage over half-duplex (HD), it is prone to self-interference (SI) due to the simultaneous transmission and reception of the transceiver (TRX).
This requires advanced SI cancellation (SIC) techniques to overcome the drawback.

Till now, SIC has been developed from different levels. Taken as a whole, SI can be suppressed by passive SIC approaches \cite{AT1,AT2}, analog/digital circuit design \cite{AS2,DS}, and beamforming \cite{BF1,BF2,BF3,BF4,BF5,BF6,BF7}. First, passive SIC approaches suppress SI in antenna-/propagation-domain, such as antenna isolation. There are two main functions served by analog SIC: 1) eliminating imperfections from the transmit RF chain, and 2) avoiding analog saturation on the receiver (RX) end.
After analog SIC, the goal of digital SIC is to reduce any remaining SI, which may include both linear and large nonlinear factors. 
For mmWave systems with massive antennas, analog SIC is not preferred since each transmit/receive antenna is equipped with a power amplifier/low noise amplifier. This will incur a large complexity for analog SIC. As the number of RF chains stays low with mmWave hybrid beamforming architectures, digital SIC stands as a viable option.

 Moreover, beamforming is a significant technique that exploits the multi-antenna spatial characteristic for SIC without any additional analog circuitry. In this sense, beamforming aims to maximize spatial multiplex gain while suppressing SI, such as null-space projection and minimum mean-squared error (MMSE) filters \cite{BF1,BF2}, and interference nulling \cite{BF3}.
 On the other hand, beamforming-based SIC for mmWave FD is different from that for sub-6G FD due to the hybrid analog and digital beamforming design \cite{HB1,HB2}. With the hybrid beamforming architecture,
  the authors in \cite{BF4} designed optimal analog beamforming for SIC but brought much performance loss due to the projection error onto the subspace of the modulus-1 constraint. 
 In \cite{BF5,BF7}, hybrid beamforming with iterative optimization was designed to minimize the SI power and improve the SE. In \cite{BF6}, hybrid beamforming with the Nystr$\ddot{\text{o}}$m method for cost-effective low-rank approximation was proposed for integrated access and backhaul with FD systems.  
 
In addition to the SI problem, FD cannot solve the propagation problem of mmWave itself, i.e., severe path loss and the easily blocked line-of-sight (LoS) path. Thanks to advancements in micro-electrical-mechanical systems and programmable metamaterials, reconfigurable intelligent surfaces (RISs), with their low power consumption, low hardware cost, and reliable ability to improve channel environments \cite{RIS1,RIS2}, deserve to be introduced into FD mmWave systems for reliability enhancement. 
As a promising technique, RIS is well suited for integration with other technologies, such as mmWave communications \cite{CS1,CS2,CS3,CS4,AMP,HP1,HP2}, simultaneous wireless information and power transfer \cite{SWIPT2,SWIPT1}, and age of information \cite{SWIPT1,AOI1,AOI2}, to produce more robust systems. Particularly, the research on RIS-aided mmWave communications is increasing, with a focus on channel estimation \cite{CS1,CS2,CS3,CS4,AMP}, and hybrid and passive beamforming \cite{HP1,HP2}. These studies considered the HD system.
The research most relevant to our work, i.e., RIS-aided FD, was documented in \cite{OP,MUFR,multi-hop,STAR,REC2}.
 Through studying the outage probability, the authors in \cite{OP} demonstrated the superiority of the RIS-aided two-way systems compared to the one-way  counterpart. Reference \cite{MUFR} showed that the RIS in FD systems could suppress the users' interference since the RIS created effective reflecting paths between the base station (BS) and users (UEs).
  In \cite{multi-hop}, multi-hop RIS-aided FD relay systems were investigated by deriving outage probability, average SE, and average bit error rate.
Moreover, the FD system with the assistance of simultaneous transmission and reflection-based RIS was investigated in \cite{STAR}.
 
 However, to the best of our knowledge, all these previous work on RIS-aided FD considered sub-6G, without mmWave hybrid beamforming architectures. Moreover, research on sparse channel estimation for the FD arrays with spatial correlation is still lacking. Motivated by these gaps, this work aims to investigate channel estimation, passive beamforming, and hybrid beamforming for RIS-aided FD mmWave communications. The main contributions are as follows:

\begin{itemize}
	\item \emph{Angular-Domain Reciprocity of FD arrays:}
A FD TRX is equipped with two antenna arrays, for respective transmission and reception, such that they are relatively near compared with a far source/scatter. Thus, some works \cite{STAR,REC1,REC2,REC3} made the assumption that the UL and DL channel matrices are reciprocal (for example, $\mathbf{H}_{{\rm D},2}=\mathbf{H}_{{\rm D},1}^H$ in Fig. \ref{system1}.). This is incompletely correct due to the phase difference between the TRX arrays (similar to the inter-element phase difference in an antenna array). To account for this, the angular-domain reciprocity property of FD arrays is demonstrated. 
	We show that, under the same scattering environment, the TRX arrays generally adhere to the far-field planar wavefront assumption, and we derive the maximal spacing between TRX arrays when the planar wavefront assumption holds.
	Owing to the planar wavefront assumption, the phase difference between the TRX arrays is a function of the inter-array spacing and  reference angles. This indicates that the TRX arrays can share the reference angle for planar array responses (PARs).

	
	\item \emph{Joint DL-UL Channel Estimation:} Leveraging on the angular-domain reciprocity property and channel sparsity, different compressive sensing (CS)-based channel estimation frameworks are proposed by jointly recovering DL-UL channel parameters.
	First, for UE's SI channel estimation\footnote{Since the BS's surrounding environment generally  keeps fixed, its SI channel can be estimated before access link establishment. Hence we do not consider BS's SI channel estimation.\label{BSI}}, we propose a Khatri-Rao product-based CS (KR-CS) framework, which performs better and is $G$ times faster than Kronecker CS (K-CS) \cite{KCS}, where $G$ is the size of dictionaries used for TRX arrays' parameter recovery. Moreover, a distributed CS (D-CS) framework, with multiple vectors sharing the sparsity but different sensing matrices, is developed for direct DL-UL channel estimation. For cascaded channel estimation, a two-stage multiple measurement vector-based D-CS (M-D-CS) framework is  proposed for BS's and UE's angle-of-arrival (AoA)/angle-of departure (AoD) estimation. Based on the cascaded structural sparsity, a parallelizable multiple $1$-sparse vector recovery problem is formulated to estimate RIS's AoA/AoD. Further, we employ Look Ahead Orthogonal Match Pursuit (LAOMP) as the basic recovery algorithm to solve the above proposed CS frameworks. Due to the basis mismatch problem \cite{mismatch} inherent in angle estimation, the off-gird refinement method is utilized for estimation performance enhancement. 
 
	\item \emph{Passive and Hybrid Beamforming:} Due to the large complexity of joint passive and hybrid beamforming for SIC, we separate them with passive beamforming maximizing the cascaded channel capacity, and hybrid beamforming maximizing the spatial multiplex gain while minimizing the SI. For this, by fixing the hybrid precoder and combiner, we demonstrate that the cascaded SE with respect to (w.r.t.) passive beamforming can be maximized with the angular-domain cascaded channel. 
 Based on the optimized passive beamforming, the weighted minimum mean squared error (WMMSE) algorithm that establishes the equivalence between	sum-rate functions and the weighted sum-MSE is employed for fully-digital SIC. Then, hybrid WMMSE for SIC (H-WMMSE-SIC) is proposed by jointly optimizing
	the hybrid precoder and combiner. In each iteration, the digital combiner/precoder can be solved by WMMSE with the analog combiner/percoder fixed, followed by a coordinate descent (CD) method for analog combiner/percoder optimization.
\end{itemize}

The rest of this paper is organized as follows.
Section \ref{Sec2} describes the RIS-aided FD communication scenario, the signal model and the channel model. Particularly, angular-domain reciprocity for FD arrays is shown. In Section \ref{Sec4}, different CS frameworks are proposed for joint DL-UL channel estimation. Section \ref{Sec5} proposes angular-domain-based passive beamforming and the H-WMMSE-SIC method. In Section \ref{Sec6}, numerical results are carried out to demonstrate the effectiveness of our proposed methods. Finally, conclusion is made in Section \ref{Sec7}.

\emph{Notations:} ${\left(  \cdot  \right)}^{ *}$, ${\left(  \cdot  \right)}^{ T}$ and ${\left(  \cdot  \right)}^{ H}$ denote conjugate, transpose, conjugate transpose, respectively.  $\Vert\cdot\Vert_0$ and $\Vert\cdot\Vert_2$ represent $\ell_0$ norm and $\ell_2$ norm, respectively. 
$\Vert\mathbf{A}\Vert_F$ denotes the Frobenius norm of matrix $\mathbf{A}$. Furthermore, $\bullet$, $\otimes$, and $\odot$ are the KR product, the Kronecker product and the Hadamard product, respectively. $[\mathbf{a}]_{i}$ and $[\mathbf{A}]_{i,j}$ denote the $i$-th element of vector $\mathbf{a}$, the $(i,j)$-th element of matrix $\mathbf{A}$, respectively. $\rm{vec}(\cdot)$ represents the vectorization operation. $\mathbb{E}\{\cdot\}$ is the expectation operator.
$\Re\{a\}$ denotes the real part of complex $a$. $\mathbf{I}_M$ denotes the $M$-by-$M$ identity matrix. Moreover, $\rm{diag}(\mathbf{a})$ is a square diagonal matrix with entries of $\mathbf{a}$ on its diagonal. Finally, $\mathcal{CN}(\mathbf{a},\mathbf{A})$ is the complex Gaussian distribution with mean $\mathbf{a}$ and covariance matrix $\mathbf{A}$.
\vspace{-0.4cm}
\section{System Model}\label{Sec2}

\begin{figure}
	\begin{minipage}[t]{0.496\linewidth}
		\centering
		\includegraphics[height=5.5cm,width=7.35cm]{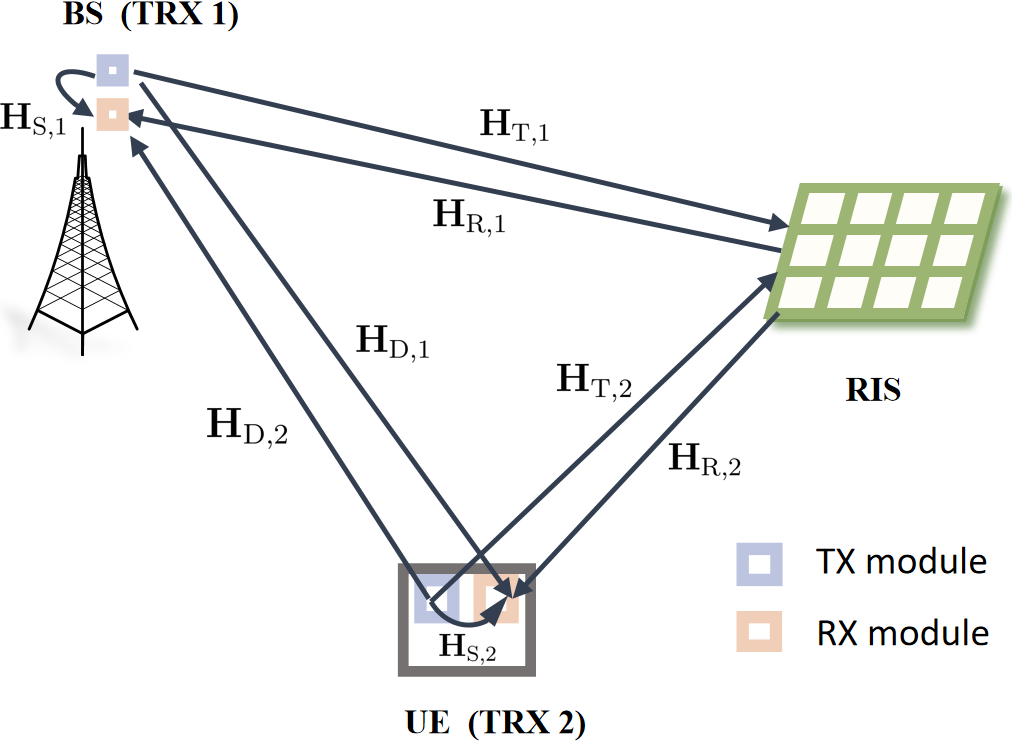}
		\caption{A RIS-aided in-band FD  system.}\label{system1}
	\end{minipage}%
	\begin{minipage}[t]{0.49\linewidth}
		\centering
		\includegraphics[height=5.5cm,width=8.22cm]{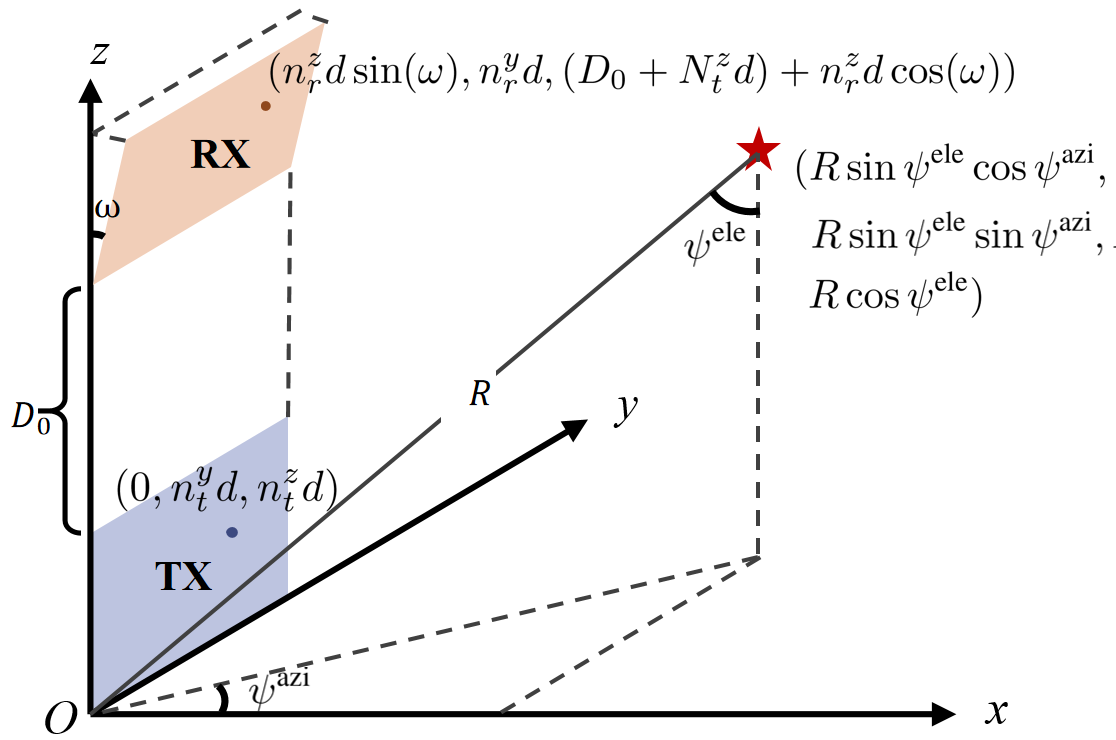}
		\caption{A structure of the 3D TRX array.}\label{system2}
	\end{minipage}
	
\end{figure}

 As shown in Fig. \ref{system1}, we consider a RIS-aided in-band FD mmWave system with a BS (TRX 1) and a UE (TRX 2), where
$\{\mathbf{H}_{{\rm D},i}\}_{i=1}^2$ are the direct channels, $\mathbf{H}_{{\rm T},i},\forall i\in\{1,2\}$, is the DL channel from TRX $i$ to the RIS, $\mathbf{H}_{{\rm R},i},\forall i\in\{1,2\}$, is the UL channel from TRX $i$ to the RIS.
Due to the simultaneous transmission and reception, as well as the in-band frequency resource reuse in FD systems, the SI is not avoidable. We denote the SI channel of TRX $i$ by $\mathbf{H}_{{\rm S},i}, \forall i\in\{1,2\}$.
 Assuming that all antenna arrays of the BS, UE, and RIS are equipped with uniform planar arrays (UPAs).
 $N_{1}\triangleq \underbrace{N_{1,t}^zN_{1,t}^y}_{N_{1,t}}+\underbrace{N_{1,r}^zN_{1,r}^y}_{N_{1,r}}$ and $N_{2}\triangleq \underbrace{N_{2,t}^zN_{2,t}^y}_{N_{2,t}}+\underbrace{N_{2,r}^zN_{2,r}^y}_{N_{2,r}}$ are the number of antenna elements of TRX 1 and TRX 2, respectively, in which the subscripts $t$/$r$ and $z/y$ denotes the TX/RX array and the $z$-/$y$-axis, respectively. The number of the RIS's reflective elements is $L$. With a passive RIS, each element has a constant modulus constraint. Denoting the RIS reflection matrix by $\mathbf{V}\triangleq{\rm diag}(\mathbf{v})\in\mathbb{C}^{L\times L}$, where $\mathbf{v}\triangleq[v_1,v_2,\cdots,v_L]\in\mathbb{C}^{1\times L}$ with $\vert v_i\vert=1$.
 $\forall i\in\{1,2\}$,
TX $i$ and RX $i$ are equipped with $M_{i,t}$ and $M_{i,r}$ RF chains, respectively. In a fully-digital system with one dedicated RF chain per antenna, each TRX follows $M_{i,t}= N_{i,t}$ and  $M_{i,r}= N_{i,r}$.
With a fully-connected hybrid beamforming architecture (each RF chain connects to all antenna elements), a small number of RF chains is considered, such that $M_{i,t}\ll N_{i,t}$ and  $M_{i,r}\ll N_{i,r}$,
 for low hardware cost and energy consumption. Besides, we denote the number of transmission data streams of TX $i$ by $N_{st,i}$, such that $N_{st,i}\leq {\rm min}\{M_{i,t},M_{i,r}\},\forall i\in\{1,2\}$.

\subsection{Signal Model}

$\mathbf{F}_i\triangleq\mathbf{F}_{{\rm RF},i}\mathbf{F}_{{\rm BB},i}\in\mathbb{C}^{N_{i,t}\times N_{s,i}}$ and $\mathbf{W}_i\triangleq\mathbf{W}_{{\rm RF},i}\mathbf{W}_{{\rm BB},i}\in\mathbb{C}^{N_{i,r}\times N_{s,i}}$ are respectively fully-digital precoder and combiner of TRX $i$, where $\mathbf{F}_{{\rm RF},i}\in\mathbb{C}^{N_{i,t}\times M_{i,t}}$, $\mathbf{F}_{{\rm BB},i}\in\mathbb{C}^{M_{i,t}\times N_{s,i}}$, $\mathbf{W}_{{\rm RF},i}\in\mathbb{C}^{N_{i,r}\times M_{i,r}}$ and $\mathbf{W}_{{\rm BB},i}\in\mathbb{C}^{M_{i,r}\times N_{s,i}}$ denote analog precoding, digital precoding, analog combiner and digital combiner, respectively.
 When TX $i$ ($\forall i\in\{1,2\}$) transmits the signal $\mathbf{s}_{i}\in\mathbb{C}^{N_{s,i}\times 1}$ with $\mathbb{E}[\mathbf{s}_i\mathbf{s}_i^H]=\frac{1}{N_{st,i}}\mathbf{I}_{N_{st,i}}$,
 the received signal at RX $j$ ($\forall j\in \{1,2\},j\neq i$), is given by 
\begin{equation}\label{Yj}
	\setlength{\abovedisplayskip}{3.2pt}
	\begin{aligned}
\mathbf{y}_j=&{\mathbf{W}^H_j\mathbf{H}_{{\rm D},i}\mathbf{F}_i\mathbf{s}_i}+{\mathbf{W}^H_j\mathbf{H}_{{\rm R},j}\mathbf{V}\mathbf{H}_{{\rm T},i}\mathbf{F}_i\mathbf{s}_i}+{\mathbf{W}^H_j\mathbf{H}_{{\rm S},j}\mathbf{F}_j\mathbf{s}_j}+\mathbf{W}_j^H\mathbf{n}\\
=&\underbrace{\mathbf{W}^H_{{\rm BB},j}\mathbf{W}^H_{{\rm RF},j}\mathbf{H}_{{\rm D},i}\mathbf{F}_{{\rm RF},i}\mathbf{F}_{{\rm BB},i}\mathbf{s}_i}_{\text{direct  signal}}
+\underbrace{\mathbf{W}^H_{{\rm BB},j}\mathbf{W}^H_{{\rm RF},j}\mathbf{H}_{{\rm R},j}\mathbf{V}\mathbf{H}_{{\rm T},i}\mathbf{F}_{{\rm RF},i}\mathbf{F}_{{\rm BB},i}\mathbf{s}_i}_{\text{cascaded  signal}}\\
&+\underbrace{\mathbf{W}^H_{{\rm BB},j}\mathbf{W}^H_{{\rm RF},j}\mathbf{H}_{{\rm S},j}\mathbf{F}_{{\rm RF},j}\mathbf{F}_{{\rm BB},j}\mathbf{s}_j}_{\text{self-interference}}+\mathbf{W}^H_{{\rm BB},j}\mathbf{W}^H_{{\rm RF},j}\mathbf{n}_j,
	\end{aligned}
\end{equation}
where $\mathbf{n}_j\in\mathbb{C}^{N_{r,j}\times 1}$ is the additive white Gaussian noise  vector following $\mathcal{CN}(0,\sigma_n^2\mathbf{I}_{N_{r,j}})$.

\vspace{-0.4cm}
\subsection{Channel Model}
We show in Fig. \ref{system2} the duplex UPA structure, from which we can see that the whole array structure depends on the relative distance $D_{0}$ and the relative angle $\omega$. Theoretically, the SI becomes smaller as the relative distance $D_{0}$ becomes larger, but practical TRXs, such as cell-phones, have physical space limitations in practice. In this sense, the spatial correlation of TRX arrays is worthy studying.
 
\subsubsection{Angular-Domain Reciprocity}
 Considering that the channels between a source and the TRX arrays have spatial correlations owing to the common scattering environment,
the angular-domain reciprocity of the TRX channels holds due to the far-field planar wavefront assumption\footnote{
In the far-field planar wavefront assumption,	all antenna elements transmit/receive signals along the same AoD/AoA to formulate the commonly used planar-wave array response (PAR). 
 }. That is, when the distance between the TRX arrays and the source is far enough, the planar wavefront is assumed to make the TRX arrays' direction angles have a constant phase difference. 

First, we denote the indices of the TX and RX array elements
by sets $\mathcal{L}_t\triangleq\{(n_t^z,n_t^y)|n_t^z=0,\cdots,N_t^z-1; n_t^y=0,\cdots,N_t^y-1\}$ and $\mathcal{L}_r\triangleq\{(n_r^z,n_r^y)|n_r^z=0,\cdots,N_r^z-1; n_r^y=0,\cdots,N_r^y-1\}$, respectively.
The two arrays' spherical array responses (SARs), without any approximation assumption, can be written as
\begin{equation}\label{bt}
\setlength{\abovedisplayskip}{3.2pt}
	\begin{aligned}
\mathbf{b}_{t}(R,\mathcal{L}_t)\triangleq\sqrt{\frac{1}{N_t}}\left[e^{j\frac{2\pi}{\lambda}\left(R-R_{t,(0,0)}\right)},\cdots,e^{j\frac{2\pi}{\lambda}\left(R-R_{t,(N_t^z,N_t^y)}\right)}\right]^T
	\end{aligned}
\end{equation}
and
\begin{equation}\label{br}
	\setlength{\abovedisplayskip}{3.2pt}
	\begin{aligned}
	\mathbf{b}_{r}(R,\mathcal{L}_r)\triangleq\sqrt{\frac{1}{N_r}}\left[e^{j\frac{2\pi}{\lambda}\left(R-R_{r,(0,0)}\right)},\cdots,e^{j\frac{2\pi}{\lambda}\left(R-R_{r,(N_r^z,N_r^y)}\right)}\right]^T,
	\end{aligned}
\end{equation}
where $R$ is the distance between the source/scatter and the array reference point. $\frac{2\pi}{\lambda}$ is the wave number with 
$\lambda$ being the wavelength.
 $R_{t,(n_t^z,n_t^y)}$ and $R_{r,(n_r^z,n_r^y)}$ represent the distance between the source/scatter and the $(n_t^z,n_t^y)$-th TX element and between the source/scatter and the $(n_r^z,n_r^y)$-th RX element, respectively.
 
 As shown in Fig. \ref{system2}, let $\psi^{\rm ele/azi}$ and $\psi^{\rm ele/azi}$ be the elevation/azimuth angles between the reference point and the source/scatter, respectively. 
 The three-dimensional (3D) Cartesian coordinates of the $(n_t^z,n_t^y)$-th TX element and 
 the $(n_r^z,n_r^y)$-th RX element are  $\left(0,y_{t,(n_t^z,n_t^y)},z_{t,(n_t^z,n_t^y)}\right)$ and $\left(0,y_{r,(n_r^z,n_r^y)},z_{r,(n_r^z,n_r^y)}\right)$, respectively.
 Assuming each array's antenna inter-element spacing is $d$, the 3D Cartesian coordinate of the $(n_t^z,n_t^y)$-th TX element is simplified to $\left(0,n^y_td,n_t^zd\right)$. Similarly, the 3D Cartesian coordinate of the $(n_r^z,n_r^y)$-th RX element is simplified to $(n_r^zd\sin(\omega),n_r^yd,(D_0+N_t^zd)+n_r^zd\cos(\omega))$. 
 Then the vector coordinates $\overrightarrow{\mathbf{R}}_{t,(n_t^z,n_t^y)}$ and $\overrightarrow{\mathbf{R}}_{r,(n_r^z,n_r^y)}$ are given by
 \begin{equation}
 	\setlength{\abovedisplayskip}{3.2pt}
 	\begin{aligned}
\overrightarrow{\mathbf{R}}_{t,(n_t^z,n_t^y)}=\left(R\sin(\psi^{\rm ele})\cos(\psi^{\rm azi}),
R\sin(\psi^{\rm ele})\sin(\psi^{\rm azi})-n_t^yd, \ R\cos(\psi^{\rm ele})-n_t^zd\right).
 	\end{aligned}
 \end{equation}
 \begin{equation}\label{Rr}
 	\setlength{\abovedisplayskip}{3.2pt}
	\begin{aligned}
	\overrightarrow{\mathbf{R}}_{r,(n_r^z,n_r^y)}=& \left(R\sin(\psi^{\rm ele})\cos(\psi^{\rm azi})-n_r^zd\sin(\omega),
		R\sin(\psi^{\rm ele})\sin(\psi^{\rm azi})-n_r^yd, \right. \\ & \left. \ R\cos(\psi^{\rm ele})-(D_0+N_t^zd)-n_r^zd\cos(\omega)\right).
	\end{aligned}
\setlength{\belowdisplayskip}{3.2pt}
\end{equation}
  Further, $R_{t,(n_t^z,n_t^y)}$ and $R_{r,(n_r^z,n_r^y)}$ are calculated by
$R_{t,(n_t^z,n_t^y)}=\left|\overrightarrow{\mathbf{R}}_{t,(n_t^z,n_t^y)}\right|$ and $R_{r,(n_r^z,n_r^y)}=\left|\overrightarrow{\mathbf{R}}_{r,(n_r^z,n_r^y)}\right|$, respectively. For clarity, denote by $\Psi^{\rm azi}\triangleq\sin(\psi^{\rm ele})\sin(\psi^{\rm azi})$ and $\Psi^{\rm ele}\triangleq\cos(\psi^{\rm ele})$, 
the SAR of Eqn. (\ref{bt}) can be approximated as the commonly used PAR in the far-field region:
\begin{equation}\label{at}
	\begin{aligned}
		\setlength{\abovedisplayskip}{3.2pt}
\mathbf{a}_t(\Psi^{\rm ele},\Psi^{\rm azi})=\sqrt{\frac{1}{N_t}}\left[1,\cdots,
e^{j\frac{2\pi}{\lambda}(n_t^yd\Psi^{\rm azi}+n_t^zd\Psi^{\rm ele})}, \cdots,
e^{j\frac{2\pi}{\lambda}\left((N_t^y-1)d\Psi^{\rm azi}+(N_t^z-1)d\Psi^{\rm ele}\right)}\right]^T,
	\end{aligned}
\setlength{\belowdisplayskip}{3.2pt}
\end{equation}
where $\forall (n_t^z,n_t^y)\in \mathcal{L}_t$.
Generally, when the phase error, w.r.t. the array aperture and distance $R$, between Eqn. (\ref{bt}) and (\ref{at}) is smaller than the maximal tolerant error, it can be ignored. This is also called the planar wavefront assumption.
 The boundary of the far- and near-field region is called the Rayleigh distance.
  
Our interest regarding the angular-domain reciprocity is whether the RX's array response (AR), as same as Eqn. (\ref{at}), can also be relevant to $\{\Psi^{\rm ele},\Psi^{\rm azi}\}$ only. However, the relative distance $D_0$ may aggravate the phase error. For example, when $D_0$ is large enough, the two arrays' direction angles are naturally different. In practice, large $D_0$ can reduce the SI but decrease the TRX channels' correlation and the space utilization. The following proposition shows the angular-domain reciprocity of the TRX channels at a specific maximal phase error.

\emph{Proposition 1: 
	Assuming that the TRX arrays in Fig. \ref{system2} have the same scattering environment,
	the TRX channels have angular-domain reciprocity when the relative distance $D_0\leq D_{\rm max}$. For simplicity and to coincide with Eqn. (\ref{at}), we set $\omega=0$ in this work\footnote{When $\omega\neq0$, the angular-domain reciprocity still holds.}.
	Let $\delta_{\rm max}$ and $R_{\rm min}$ be the maximal phase error and
	the distance between the reference point and the closest scatter, respectively, the maximal relative distance is given by $D_{\rm max}=\sqrt{\frac{\lambda}{\pi}R_{\rm min}\delta_{\rm max}-(N_r^yd)^2}-N_t^zd-N_r^zd$.}

\emph{Proof: See Appendix \ref{appendixA}.} 

Considering a system with $D=20\lambda$, $\delta_{\rm max}=\frac{\pi}{4}$, $N_r^y=N_r^z=N_t^z=N_t^y=8$, $\lambda=3$ mm, and $d=\frac{\lambda}{2}$, the Rayleigh distance is $9.6$ m. That is, in the case when the communication distance exceeds  $9.6$ m, the TRX arrays have the angular-domain reciprocity property; otherwise, the reciprocity property will not be related to the angle but also the distance.
We assume that $D_0\leq D_{\rm max}$ always holds, i.e., the angular-domain reciprocity is considered in this work.

According to the above proposition, the RX's PAR at $\omega=0$ can be written as
\begin{equation}\label{ar}
	\begin{aligned}
	\mathbf{a}_r(\Psi^{\rm ele},\Psi^{\rm azi})&=\sqrt{\frac{1}{N_r}}\left[e^{j\frac{2\pi}{\lambda}D_t\Psi^{\rm ele}},\cdots,e^{j\frac{2\pi}{\lambda}\left(n_r^yd\Psi^{\rm azi}+(D_t+n_r^zd)\Psi^{\rm ele}\right)}   ,\cdots,
	e^{j\frac{2\pi}{\lambda}\left((N_r^y-1)d\Psi^{\rm azi}+(D_t+(N_r^z-1)d)\Psi^{\rm ele}\right)}\right]^T 
	\\&=\sqrt{\frac{1}{N_r}}e^{j\frac{2\pi}{\lambda}D_t\Psi^{\rm ele}}\left[1,\cdots,e^{j\frac{2\pi}{\lambda}\left(n_r^yd\Psi^{\rm azi}+n_r^zd\Psi^{\rm ele}\right)}   ,\cdots,
	e^{j\frac{2\pi}{\lambda}\left((N_r^y-1)d\Psi^{\rm azi}+(N_r^z-1)d\Psi^{\rm ele}\right)}\right]^T,
		\end{aligned}
	\setlength{\belowdisplayskip}{3.2pt}
\end{equation}
where $D_t=D_0+(N_t^z-1)d$, and $\forall (n_r^z,n_r^y)\in \mathcal{L}_r$.
\subsubsection{Direct and Cascaded Channel Model} 
Due to the limited number of scatters in mmWave channels, the channel can be expressed as the sum of the outer product of ARs with limited paths. According to the angular-domain reciprocity, the TX $i$'s AoDs are identical to the RX $i$'s AoAs, and the number of paths between the uplink and downlink channels, denoted by $A$, is the same.
Hence the direct channel $\mathbf{H}_{{\rm D},i} (\forall i\in\{1,2\})$ can be given by
\begin{equation}
\mathbf{H}_{{\rm D},i}=\sqrt{\frac{N_{i,t}N_{i,r}}{A}}\sum_{a=1}^A \alpha_{i,a}\mathbf{a}_{r,j}(\theta_{j,a}^{\rm ele},\theta_{j,a}^{\rm azi})\mathbf{a}_{t,i}^H(\theta^{\rm ele}_{i,a},\theta^{\rm azi}_{i,a}),
\end{equation}
where $\forall j=\{1,2\}, j\neq i$. $\theta^{\rm ele/azi}_{1,a}$ and $\theta^{\rm ele/azi}_{2,a}$ denote the $a$-th path's virtual direction angles of the TRX 1 and 2, respectively, and $\alpha_{1,a}$ and $\alpha_{2,a}$ are their corresponding path gains. Besides, $\mathbf{a}_{t,1}$ and $\mathbf{a}_{t,2}$ have a similar definition as Eqn. (\ref{at}), but with different antenna elements ($N_{1,t}$ and $N_{2,t}$). The following definitions regarding the AR's numeric subscript in this paper are all like this.
Similarly, let $B_1$ be the number of paths of $\{\mathbf{H}_{{\rm T},1},\mathbf{H}_{{\rm R},1}\}$, and $B_2$ be the number of paths of $\{\mathbf{H}_{{\rm T},2},\mathbf{H}_{{\rm R},2}\}$.
The channels are expressed as
\begin{equation}
	\setlength{\abovedisplayskip}{3.2pt}
	\mathbf{H}_{{\rm T},i}=\sqrt{\frac{N_{i,t}L}{B_i}}\sum_{b_i=1}^{B_i} \beta_{i,b_i} \mathbf{a}_{l}(\varrho_{i,b_i}^{\rm ele},\varrho_{i,b_i}^{\rm azi})\mathbf{a}_{t,i}^H(\vartheta^{\rm ele}_{i,b_i},\vartheta^{\rm azi}_{i,b_i}),
\end{equation}
\begin{equation}
	\setlength{\abovedisplayskip}{3.2pt}
	\mathbf{H}_{{\rm R},i}=\sqrt{\frac{N_{i,r}L}{B_i}}\sum_{b_i=1}^{B_i} \chi_{i,b_i} \mathbf{a}_{r,i}(\vartheta^{\rm ele}_{i,b_i},\vartheta^{\rm azi}_{i,b_i})\mathbf{a}_{l}^H(\varrho_{i,b_i}^{\rm ele},\varrho_{i,b_i}^{\rm azi}),
\end{equation}
where $\mathbf{a}_l$ is the PAR of the RIS.
\subsubsection{SI Channel} Since the two arrays are nearby, the LoS path of the channel between them cannot satisfy the planar wavefront assumption. Therefore,
 we can recognize that the SI channel consists of not only the SAR for the LoS path, but also the PAR for NLoS paths. The SI channel comprised of the LoS component
is given by
\begin{equation}\label{SLOS}
	\setlength{\abovedisplayskip}{3.2pt}
\mathbf{H}_{{\rm S},i}^{\rm LoS}=\sqrt{N_{i,t}N_{i,r}}\gamma_i^{\rm LoS}\mathbf{b}_{r,i}(R_{s_i},\mathcal{L}_{r,i})\mathbf{b}^H_{t,i}(R_{s_i},\mathcal{L}_{t,i}),
\setlength{\belowdisplayskip}{3.2pt}
\end{equation}
where $i\in\{1,2\}$, $R_{s_i}$ denotes the distance between the TX $i$ and RX $i$, and $\gamma_i^{\rm LoS}$ is the LoS path gain.
Considering the NLoS channel, the identical AoA and AoD are used for TX/RX PARs owing to the angular-domain reciprocity. Thus, we have
\begin{equation}\label{SNLOS}
	\begin{aligned}
	\mathbf{H}_{{\rm S},i}^{\rm NLoS}= \sqrt{\frac{N_{i,t}N_{i,r}}{P_i}}\sum_{p_i=1}^{P_i}\gamma^{\rm NLoS}_{p_i}\mathbf{a}_{r,i}(\phi_{i,p_i}^{\rm ele},\phi_{i,p_i}^{\rm azi})\mathbf{a}^H_{t,i}(\phi_{i,p_i}^{\rm ele},\phi_{i,p_i}^{\rm azi}),
	\end{aligned}
\setlength{\belowdisplayskip}{3.2pt}
\end{equation}
where $P_i$ is the number of paths of TRX $i$'s NLoS channel, $\gamma^{\rm NLoS}_{p_i}$ is the path gain, $\phi^{\rm ele/azi}_{i,p_i}$ is the direction angle from the reference point to the $p_i$-th scatter. 
Then the whole SI channel is given by
$\mathbf{H}_{{\rm S},i}=\mathbf{H}_{{\rm S},i}^{\rm LoS}+	\mathbf{H}_{{\rm S},i}^{\rm NLoS}$.

\subsubsection{Virtual Channel Representation (VCR)}\label{VCR}
The key of sparse channel estimation is to virtually represent the true channel by the pre-defined angular-domain dictionary. Generally, the widely used dictionary for sparse channel estimation is the over-complete discrete Fourier transform (DFT) dictionary, which can effectively characterize the beamspace. More precisely, for the standard PAR of UPAs, such as Eqn. (\ref{at}), the over-complete DFT dictionary is formulated by sampling $\Psi^{\rm ele/azi}$ in a over-complete DFT bin: $\{-1,-1+\frac{2}{G^{z/y}},\cdots,1-\frac{2}{G^{z/y}}\}$, where $G^z$ and $G^y$ denote the sample points along the elevation and azimuth directions, respectively. $G=G^y\times G^z$ is the total sample points over the 3D beamspace.
 Therefore, the over-complete DFT dictionary w.r.t. the TX array is given by $\mathbf{A}_{\rm T}$, with the atom being $\mathbf{a}_t\left(-1+\frac{2g^z}{G^z},-1+\frac{2g^y}{G^y}\right)$, $\forall (g^z,g^y)\in \mathcal{G}\triangleq\{(g^z,g^y)|g^z=0,1,\cdots,G^z;g^y=0,1,\cdots,G^y\}$. 
 Besides, the RX array's dictionary can be represented by the same sample angles used for the TX array's dictionary, i.e., $\mathbf{A}_{\rm R}$ with the atom being $\mathbf{a}_r\left(-1+\frac{2g^z}{G^z},-1+\frac{2g^y}{G^y}\right)$, $\forall (g^z,g^y)\in \mathcal{G}\triangleq\{(g^z,g^y)|g^z=0,1,\cdots,G^z;g^y=0,1,\cdots,G^y\}$. 
  Taking the VCR of $\mathbf{H}_{{\rm D},1}$ as an example, we have
$
\mathbf{H}_{{\rm D},1}\approx\mathbf{A}_{{\rm R},2}\bm{\mathcal{A}}_1\mathbf{A}_{{\rm T},1}^H$, where
 $\bm{\mathcal{A}}_1\in\mathbb{C}^{G\times G}$ is a sparse matrix, the non-zero element of which indicates the AoD, AoA and path gain information. 
 Similarly, the other channels can also be virtually represented by over-complete DFT dictionaries. 

\section{Joint DL-UL Channel Estimation}\label{Sec4}
The whole channel estimation procedure consists of two phases.
In the first phase, the RIS is turned off for easy estimation of the SI channel and the direct channel. 
This work takes the SI channel of the BS as $\emph{a-priori}$ information\textsuperscript{\ref{BSI}}, and focuses on the estimation of the UE's SI channel. In this sense, only TX 2/RX 2 transmits/receives pilot signals. 
After the SI channel is estimated, the downlink and uplink channels are jointly estimated, owing to the FD mode and the angular-domain reciprocity, with a low training complexity.
In the second phase, the cascaded channel parameters are estimated. Since there are various parameters (the BS's, UE's and RIS's direction angles) required to be estimated, it will result in an unacceptable pilot overhead if using conventional RIS-based channel estimation methods \cite{CS1,CS2}. Therefore, we propose a joint UL-DL cascaded channel estimation approach, which first estimates the BS's and UE' direction angles at a fixed RIS phase matrix, and then estimates the RIS's direction angles based on the estimated BS's and UE' direction angles.
Particularly, this phase leverages on the angular-domain reciprocity to better recover channel parameters. 
\vspace{-0.4cm}
\subsection{SI Channel Estimation }
As described before, the UE's SI channel varies as the scattering environment changes. However, its LoS component is slow-varing since the LoS path of the SI channel is independent of the scattering environment. According to Eqn. (\ref{SLOS}), the LoS path's SAR is easily obtained due to the known $R_{s_2}$ and $\mathcal{L}_{t/r,2}$. Hence, we can estimate its path gain $\widehat{\gamma}^{\rm LoS}_2$ by transmitting and receiving the pilot signal with the obtained SAR. Following that, the LoS component of the UE's SI channel can be reconstructed according to Eqn. (\ref{SLOS}), which is denoted by $\widehat{\mathbf{H}}_{{\rm S},2}^{\rm LoS}$.
Next, we consider $\mathbf{H}_{{\rm S},2}^{\rm NLoS}$ that is $P_2$-sparse.  Using the VCR, $\mathbf{H}_{{\rm S},2}^{\rm NLoS}$ is approximated by
$
\mathbf{H}_{{\rm S},2}^{\rm NLoS}\approx \mathbf{A}_{{\rm R},2}\bm{\Upsilon}_2\mathbf{A}_{{\rm T},2}^H
$.
Owing to the identical AoAs and AoDs of $\mathbf{H}_{{\rm S},2}^{\rm NLoS}$, which means that the row index and column index of non-zero elements in $\bm{\Upsilon}_2$ are equal, such that $\bm{\Upsilon}_2$ is a diagonal matrix. Based on this,
we develop a simplified CS solution leveraging on the Khatri-Rao product.

Assuming TX 2 sends $N_{S,X}$  pilot sequences $\mathbf{X}_{\rm S}=\left[\mathbf{F}_{2,1}\mathbf{s}_{2,1},\mathbf{F}_{2,2}\mathbf{s}_{2,2},\cdots,\mathbf{F}_{2,N_{S,X}}\mathbf{s}_{2,N_{S,X}}\right]\in\mathbb{C}^{N_{2,t}\times N_{S,X}}$ to estimate the SI channel. For each transmit pilot sequence, RX 2 spends $\frac{N_{S,Y}}{M_{2,r}}$ time slots receiving with $\widetilde{\mathbf{W}}_{\rm S}=[\mathbf{W}_{2,1},\mathbf{W}_{2,2},\cdots,\mathbf{W}_{2,\frac{N_{S,Y}}{M_{2,r}}}]\in\mathbb{C}^{N_{2,r}\times N_{S,Y}}$. After $N_{S,X}\frac{N_{S,Y}}{M_{2,r}}$ time slots, the collected signal at RX 2 can be written as
\begin{equation}
	\setlength{\abovedisplayskip}{3.2pt}
	\begin{aligned}
\mathbf{Y}_{\rm S,2}=& 
\widetilde{\mathbf{W}}_{\rm S}^H\mathbf{H}_{{\rm S},2}^{\rm NLoS}\mathbf{X}_{\rm S}+\widetilde{\mathbf{W}}_S^H\mathbf{H}_{{\rm S},2}^{\rm LoS}\mathbf{X}_{\rm S}+\mathbf{N}_{\rm S}\\
\approx&\widetilde{\mathbf{W}}_{\rm S}^H\mathbf{A}_{{\rm R},2}\bm{\Upsilon}_2\mathbf{A}_{{\rm T},2}^H\mathbf{X}_{\rm S}+\widetilde{\mathbf{W}}_{\rm S}^H\mathbf{H}_{{\rm S},2}^{\rm LoS}\mathbf{X}_{\rm S}+\mathbf{N}_{\rm S},
	\end{aligned}
\setlength{\belowdisplayskip}{3.2pt}
\end{equation}
where $\mathbf{N}_{\rm S}\in\mathbb{C}^{N_{S,Y}\times N_{S,X}}$ is the combined noise matrix.

As the previous statement, $\mathbf{H}_{{\rm S},2}^{\rm LoS}$ can be easily to be got, we assume it has been estimated to obtain
\begin{equation}\label{YS2}
	\setlength{\abovedisplayskip}{3.2pt}
\widetilde{\mathbf{Y}}_{{\rm S},2}\approx\bm{\Phi}_{\rm S, W}{\bm{\Upsilon}}_2
\bm{\Phi}_{\rm S, F}+\mathbf{N}_{\rm S},
\setlength{\belowdisplayskip}{3.2pt}
\end{equation}
where $\widetilde{\mathbf{Y}}_{{\rm S},2}=\mathbf{Y}_{\rm S,2}-\widetilde{\mathbf{W}}_{\rm S}^H\widehat{\mathbf{H}}_{{\rm S},2}^{\rm LoS}\mathbf{X}_{\rm S}$, $\bm{\Phi}_{{\rm S},{\rm W}}\triangleq\widetilde{\mathbf{W}}_{\rm S}^H\mathbf{A}_{{\rm R},2}\in\mathbb{C}^{N_{S,Y}\times G}$, and $\bm{\Phi}_{{\rm S},{\rm F}}\triangleq\mathbf{A}_{{\rm T},2}^H\mathbf{X}_{\rm S}\in\mathbb{C}^{ G\times N_{S,X}}$. Since both TX and RX PARs depend on the same angle, as shown in Eqn. (\ref{SNLOS}), it can be observed that $\bm{\Upsilon}_2$ is a diagonal matrix. 
Therefore, we propose a KR-CS framework for this special case.
 Vectorizing the above equation based on the fact that $\widetilde{\bm{\Upsilon}}_2$ is a diagonal matrix, we have
$
{\rm vec}\left(\widetilde{\mathbf{Y}}_{{\rm S},2}\right)\overset{(a)}{\approx}\left(\bm{\Phi}_{{\rm S},{\rm F}}^T\bullet\bm{\Phi}_{{\rm S},{\rm W}}\right)\bm{\upsilon}+{\rm vec}\left(\mathbf{N}_{\rm S}\right)
$,
where $(a)$ holds because ${\rm vec}(\mathbf{ABC})=(\mathbf{C}^T\bullet\mathbf{A})\mathbf{b}$ with $\mathbf{b}$ denoting the vector consisting of the diagonal elements of $\mathbf{B}$. Thus, $\bm{\upsilon}$ is a sparse vector consisting of the diagonal elements of $\widetilde{\bm{\Upsilon}}_2$. Therefore, a CS problem is formulated as
\begin{equation}\label{S2_CS}
	\setlength{\abovedisplayskip}{3.2pt}
	\begin{aligned}
& \underset{{\bm{\upsilon}}}{\rm arg \ min} \ \left\Vert{\bm{\upsilon}}\right\Vert_0 , \ {\rm s.t.}
\left\Vert{\rm vec}\left(\widetilde{\mathbf{Y}}_{{\rm S},2}\right)-\left(\bm{\Phi}_{{\rm S},{\rm F}}^T\bullet\bm{\Phi}_{{\rm S},{\rm W}}\right)\bm{\upsilon}\right\Vert_2^2\leq\epsilon,
	\end{aligned}
\setlength{\belowdisplayskip}{3.2pt}
\end{equation}
where $\epsilon$ is the precise factor.
Faced with the above problem, a variety of recovery algorithms can be applied for solving ${\bm{\upsilon}}$, e.g., greedy algorithms, convex methods and Bayesian learning. In this work, we choose LAOMP, which outperforms OMP due to the $\mathcal{I}$ loops for the look-ahead procedure \cite{LAOMP}, as the basic algorithm for its easy extensibility to various CS frameworks. 

When $\widehat{\bm{\upsilon}}$ is estimated, we can reconstruct the SI channel according to its non-zero elements. Specifically, the non-zero index indicates the AoA/AoD index in the dictionary and the non-zero value indicates the estimated path gain. Hence the reconstructed channel can be represented by $\widehat{\mathbf{H}}_{{\rm S},2}^{\rm NLOS}=\mathbf{A}_{{\rm R},2}{\rm diag}(\bm{\upsilon})\mathbf{A}_{{\rm T},2}^H$.
\vspace{-0.4cm}
\subsection{Direct Channel Estimation}
Same as SI channel estimation, we assume that the RIS is turned off.
 TX $i$ ($\forall i\in\{1,2\}$) transmits pilot sequences $\mathbf{X}_{{\rm D}_i}\triangleq[\mathbf{F}_{i,1},\mathbf{F}_{i,2},\cdots,\mathbf{F}_{i,N_{D_i,X}}]\in\mathbb{C}^{N_{i,t}\times N_{D_i,X}}$, and then for each pilot sequence, RX $j$ ($\forall j\in\{1,2\}$, $j\neq i$) receives with $\widetilde{\mathbf{W}}_{{\rm D}_j}\triangleq\left[\mathbf{W}_{j,1},\mathbf{W}_{j,2},\cdots,\mathbf{W}_{j,\frac{N_{D_j,Y}}{M_{j,r}}}\right]\in\mathbb{C}^{N_{j,r}\times N_{{D}_j},Y}$. Through $N_{{D}_i,X}\frac{N_{D_j,Y}}{M_{j,r}}$ time slots, the collected signal at RX $j$ can be given by
\begin{equation}
	\setlength{\abovedisplayskip}{3.2pt}
\mathbf{Y}_{{\rm D},j}=\widetilde{\mathbf{W}}^H_{{\rm D}_j}\mathbf{H}_{{\rm D},i}\mathbf{X}_{{\rm D}_i}+{\widetilde{\mathbf{W}}^H_{{\rm D}_j}\mathbf{H}_{{\rm S},j}\mathbf{X}_{{\rm D}_j}}+\mathbf{N}_{{\rm D}_j},
\setlength{\belowdisplayskip}{3.2pt}
\end{equation}
where $\mathbf{N}_{{\rm D}_j}\in\mathbb{C}^{N_{D_j,Y}\times N_{D_i,X}}$ is the noise matrix. 
Suppose the estimated SI channel is approximately accurate, the above equation with pilot removed is simplified as
\begin{equation}
	\setlength{\abovedisplayskip}{3.2pt}
	\begin{aligned}
\widetilde{\mathbf{Y}}_{{\rm D},j}\approx\widetilde{\mathbf{W}}^H_{{\rm D}_j}\mathbf{H}_{{\rm D},i}\mathbf{X}_{{\rm D}_i}+\mathbf{N}_{{\rm D}_j}
\approx \widetilde{\mathbf{W}}^H_{{\rm D}_j}{\mathbf{A}}_{{\rm R},j}\bm{\mathcal{A}}_i\mathbf{A}_{{\rm T},i}^H\mathbf{X}_{{\rm D}_i} +\mathbf{N}_{{\rm D}_j}
	\end{aligned}
\setlength{\belowdisplayskip}{3.2pt}
\end{equation}
where  $\bm{\mathcal{A}}_i\in\mathbb{C}^{G\times G}$ is the sparse matrix.

Leveraging on the angular-domain reciprocity, it is easy to know that the non-zero indices of $\bm{\mathcal{A}}_1$ and $\bm{\mathcal{A}}_2$ are transposed, i.e., the non-zero element's column/row index of $\bm{\mathcal{A}}_1$ is equal to the non-zero element's the row/column index of $\bm{\mathcal{A}}_2$. 
By defining 
$\bm{\Phi}_{{\rm D}_i,{\rm F}}\triangleq\mathbf{A}_{{\rm T},i}^H\mathbf{X}_{{\rm D}_i}\in\mathbb{C}^{G\times N_{D_i,X}}$ and $\bm{\Phi}_{{\rm D}_j,{\rm W}}\triangleq\widetilde{\mathbf{W}}^H_{{\rm D}_j}{\mathbf{A}}_{{\rm R},j}\in\mathbb{C}^{N_{D_j,Y}\times G}$ ($\forall i,j\in\{1,2\}$), vectorizing $\widetilde{\mathbf{Y}}_{{\rm D},1}$ and $\widetilde{\mathbf{Y}}_{{\rm D},2}^H$ yields 
$
\widetilde{\mathbf{y}}_{{\rm D},1}\triangleq{\rm vec}\left(\widetilde{\mathbf{Y}}_{{\rm D},1}\right)=\left(\bm{\Phi}_{{\rm D}_2,{\rm F}}^T\otimes \bm{\Phi}_{{\rm D}_1,{\rm W}}\right)\bm{\xi}_2+{\rm vec}\left(\mathbf{N}_{{\rm D}_1}\right)
$, and
$
\widetilde{\mathbf{y}}_{{\rm D},2}\triangleq	{\rm vec}\left(\widetilde{\mathbf{Y}}^H_{{\rm D},2}\right)=\left(\bm{\Phi}_{{\rm D}_2,{\rm W}}^*\otimes \bm{\Phi}_{{\rm D}_1,{\rm F}}^H\right)\bm{\xi}_1+{\rm vec}(\mathbf{N}_{{\rm D}_2})
$,
where $\bm{\xi}_2\triangleq{\rm vec}\left(\bm{\mathcal{A}}_2\right)$ and $\bm{\xi}_1\triangleq{\rm vec}\left(\bm{\mathcal{A}}_1^H\right)$. 
Since $\bm{\xi}_1$ and $\bm{\xi}_2$ have the same sparsity support, i.e., ${\rm supp}(\bm{\xi}_1)={\rm supp}(\bm{\xi}_2)$, the  D-CS problem is formulated:
 \begin{equation}\label{D_CS}
 	\begin{aligned}
& \ \ \  \underset{\bm{\xi}_1,\bm{\xi}_2}{\rm arg \ min} \ \left\Vert \bm{\xi}_1\right\Vert_0+\left\Vert\bm{\xi}_2\right\Vert_0 \\ {\rm s.t.} \ & 
\left\Vert\widetilde{\mathbf{y}}_{{\rm D},1}-
\left(\bm{\Phi}_{{\rm D}_2,{\rm F}}^T\otimes \bm{\Phi}_{{\rm D}_1,{\rm W}}\right)\bm{\xi}_2 \right\Vert_2^2\leq \epsilon, \\
&  \left\Vert\widetilde{\mathbf{y}}_{{\rm D},2}-
\left(\bm{\Phi}_{{\rm D}_2,{\rm W}}^*\otimes \bm{\Phi}_{{\rm D}_1,{\rm F}}^H\right)\bm{\xi}_1 \right\Vert_2^2\leq \epsilon, \\
&\ \ \ \ \ \ {\rm supp}(\bm{\xi}_1)={\rm supp}(\bm{\xi}_2).
 	\end{aligned}
 \end{equation}
To solve it, a D-LAOMP algorithm is proposed. The difference between it and LAOMP is that the former recovers multiple sparse vectors jointly with different sensing matrices. Hence D-LAOMP is comprised of multiple LAOMP procedures with a common sparsity constraint. 
Due to the space constraint, we do not describe D-LAOMP in detail.
\vspace{-0.4cm}
\subsection{Cascaded Channel Parameter Estimation}
As earlier mentioned, the cascaded channel estimation problem is divided into two subproblems: 1) BS's and UE's channel parameter estimation with a two-stage D-M-CS framework, and 2) RIS's channel parameter estimation with a parallelizable multiple $1$-sparse vector recovery framework.
\subsubsection{BS's and UE's AoA/AoD Recovery}
Since we have estimated the SI channel and direct channel, so we just consider the cascaded signal as Eqn. (\ref{Yj}). 
With a fixed RIS phase matrix, TX $i$ ($\forall i\in\{1,2\}$) transmits pilot sequences $\mathbf{X}_{{\rm C}_i}\triangleq[\mathbf{F}_{i,1},\mathbf{F}_{i,2},\cdots,\mathbf{F}_{i,N_{C_i,X}}]\in\mathbb{C}^{N_{i,t}\times N_{C_i,X}}$, and then for each pilot sequence, RX $j$ ($\forall j\in\{1,2\}$, $j\neq i$) receives with $\widetilde{\mathbf{W}}_{{\rm C}_j}\triangleq\left[\mathbf{W}_{j,1},\mathbf{W}_{j,2},\cdots,\mathbf{W}_{j,\frac{N_{C_j,Y}}{M_{j,r}}}\right]\in\mathbb{C}^{N_{j,r}\times N_{{C}_j},Y}$. With $N_{{C}_i,X}\frac{N_{C_j,Y}}{M_{j,r}}$ time slots, the residual signal $\widetilde{\mathbf{Y}}_{{\rm C},j}\in\mathbb{C}^{N_{{C}_j,Y}\times N_{{C}_i,X}}$ at RX $j$ can be given by  
\begin{equation}\label{YCJ}
	\setlength{\abovedisplayskip}{3.2pt}
	\begin{aligned}
	\widetilde{\mathbf{Y}}_{{\rm C},j} =&\widetilde{\mathbf{W}}^H_{{\rm C}_j}\mathbf{H}_{{\rm R},j}\mathbf{V}\mathbf{H}_{{\rm T},i}\mathbf{X}_{{\rm C}_i}+\mathbf{N}_{{\rm C}_j}
	\\ \overset{(b)}{=}&  \widetilde{\mathbf{W}}^H_{{\rm C}_j}\bm{\Pi}_{{\rm R},j} {\rm diag}(\bm{\chi}_j)\bm{\Pi}_{{\rm L},j}^H\mathbf{V}\bm{\Pi}_{{\rm L},i}{\rm diag}(\bm{\beta}_i)\bm{\Pi}_{{\rm T},i}^H\mathbf{X}_{{\rm C}_i}+\mathbf{N}_{{\rm C}_j}\\
	=& \widetilde{\mathbf{W}}^H_{{\rm C}_j}\bm{\Pi}_{{\rm R},j} \bm{\Lambda}_{i,j}\bm{\Pi}_{{\rm T},i}^H\mathbf{X}_{{\rm C}_i}+\mathbf{N}_{{\rm C}_j},
	\end{aligned}
\setlength{\belowdisplayskip}{3.2pt}
\end{equation}
where 
\begin{equation}\label{LAM}
	\setlength{\abovedisplayskip}{3.2pt}
\bm{\Lambda}_{i,j}\triangleq {\rm diag}(\bm{\chi}_j)\bm{\Pi}_{{\rm L},j}^H\mathbf{V}\bm{\Pi}_{{\rm L},i}{\rm diag}(\bm{\beta}_i)\in\mathbb{C}^{B_j\times B_i}
\setlength{\belowdisplayskip}{3.2pt}
\end{equation}
is the effective cascaded channel,
$\widetilde{\mathbf{Y}}_{{\rm C},j}=\mathbf{Y}_{{\rm C},j}-{\widetilde{\mathbf{W}}^H_{{\rm C}_j}\widehat{\mathbf{H}}_{{\rm D},i}\mathbf{X}_{{\rm C}_i}}-{\widetilde{\mathbf{W}}^H_{{\rm C}_j}\widehat{\mathbf{H}}_{{\rm S},j}\mathbf{X}_{{\rm C}_j}}$ with $\mathbf{Y}_{{\rm C},j}\in\mathbb{C}^{N_{{C}_j,Y}\times N_{{C}_i,X}}$ denoting the collected signal at RX $j$ in $N_{{C}_i,X}N_{C_j,Y}$ time slots. $(b)$ holds due to  $\mathbf{H}_{{\rm R},j}\triangleq\bm{\Pi}_{{\rm R},j} {\rm diag}(\bm{\chi}_j)\bm{\Pi}_{{\rm L},j}^H\in\mathbb{C}^{N_{j,r}\times L}$
and $\mathbf{H}_{{\rm T},i}\triangleq\bm{\Pi}_{{\rm L},i}{\rm diag}(\bm{\beta}_i)\bm{\Pi}_{{\rm T},i}^H\in\mathbb{C}^{ L\times N_{i,t}}$
, where $\bm{\Pi}_{{\rm R},j}\triangleq\left[\mathbf{a}_{r,i}\left(\varphi_{i,1}^{\rm ele},\varphi_{i,1}^{\rm azi}\right),\cdots,\mathbf{a}_{r,j}\left(\varphi_{j,B_j}^{\rm ele},\varphi_{j,B_j}^{\rm azi}\right)\right]\in\mathbb{C}^{N_{j,r}\times B_j}
$, $\bm{\chi}_j\triangleq\left[\chi_{j,1},\cdots,\chi_{j,B_j}\right]\in\mathbb{C}^{1\times B_j}$, $\bm{\Pi}_{{\rm L},j}\triangleq\left[\mathbf{a}_{l}\left(\varrho_{j,1}^{\rm ele},\varrho_{j,1}^{\rm azi}\right),\cdots,\mathbf{a}_{l}\left(\varrho_{j,B_j}^{\rm ele},\varrho_{j,B_j}^{\rm azi}\right)\right]\in\mathbb{C}^{L\times B_j}
$, $\bm{\Pi}_{{\rm L},i}\triangleq[\mathbf{a}_{l}  (\varrho_{i,1}^{\rm ele},\varrho_{i,1}^{\rm azi}),\cdots, \mathbf{a}_{l}(\varrho_{i,B_i}^{\rm ele},\varrho_{i,B_i}^{\rm azi})]  \in\mathbb{C}^{L\times B_i}
$, $\bm{\beta}_{i}\triangleq[\beta_{i,1},\cdots,\beta_{i,B_i}]\in\mathbb{C}^{1\times B_i}$, $\bm{\Pi}_{{\rm T},i}\triangleq[\mathbf{a}_{t,i}(\vartheta_{i,1}^{\rm ele},\vartheta_{i,1}^{\rm azi}),\cdots,\mathbf{a}_{t,i}(\vartheta_{i,B_i}^{\rm ele},\vartheta_{i,B_i}^{\rm azi})]\in\mathbb{C}^{N_{t,i}\times B_i}
$, and $\mathbf{N}_{C_j}\in\mathbb{C}^{N_{{C}_j,Y}\times N_{{C}_i,X}}$ is the noise matrix.

Based on the VCR, we approximate Eqn. (\ref{YCJ}) as $\widetilde{\mathbf{Y}}_{{\rm C},j}\approx \widetilde{\mathbf{W}}^H_{{\rm C}_j}\mathbf{A}_{{\rm R},j}
\bm{\Gamma}_{i,j}\mathbf{A}_{{\rm T},i}^H\mathbf{X}_{{\rm C}_i}+\mathbf{N}_{{\rm C}_j}$,
where $\bm{\Gamma}_{i,j}\in\mathbb{C}^{G\times G}$ is a $B_iB_j$-sparse matrix we need to recover. Note that the sparsity of $\bm{\Gamma}_{i,j}$ is structural rather than random. In $\bm{\Gamma}_{i,j}$, one non-zero row maps to $B_j$ non-zero columns, and vice versa, one non-zero column maps $B_i$ non-zero rows. This structural sparsity is called as row-column-block sparsity. Besides, $\bm{\Gamma}_{2,1}=\bm{\Gamma}_{1,2}^H$ due to the angular-domain reciprocity.
Considering the above, we propose a two-stage D-M-CS framework to recover $\bm{\Gamma}_{2,1}$.

In the first stage, we exploit the row-block-sparsity and the angular-domain reciprocity to recover the non-zero rows of $\bm{\Gamma}_{2,1}/\bm{\Gamma}_{1,2}^H$ with a D-M-CS problem:
\begin{equation}\label{dmcs1}
	\setlength{\abovedisplayskip}{3.2pt}
	\begin{aligned}
	& \ \ \ 	\underset{\mathbf{Z}_{2,1},\mathbf{Z}_{1,2}}{\rm arg \ min} \ \left\Vert \mathbf{Z}_{2,1}\right\Vert_0+\left\Vert \mathbf{Z}_{1,2}\right\Vert_0\\
	& {\rm s.t.}\ \left\Vert\widetilde{\mathbf{Y}}_{{\rm C},1}-\widetilde{\mathbf{W}}^H_{{\rm C}_1}\mathbf{A}_{{\rm R},1}\mathbf{Z}_{2,1}\right\Vert_F^2\leq \epsilon,
 \\
&\ \ \ \ \ \left\Vert\widetilde{\mathbf{Y}}_{{\rm C},2}^H-\widetilde{\mathbf{X}}^H_{{\rm C}_1}\mathbf{A}_{{\rm T},1}\mathbf{Z}_{1,2}^H\right\Vert_F^2\leq \epsilon,\\
&\ \ \ \ \
\text{ r-supp}( \mathbf{Z}_{2,1})=\text{c-supp}(\mathbf{Z}_{1,2}),
	\end{aligned}
\setlength{\belowdisplayskip}{3.2pt}
\end{equation}
where $\mathbf{Z}_{2,1}\triangleq\bm{\Gamma}_{2,1}\mathbf{A}_{{\rm T},2}^H\mathbf{X}_{{\rm C}_2}\in\mathbb{C}^{G\times N_{C_2,X}}$, $\mathbf{Z}_{1,2}\triangleq\widetilde{\mathbf{W}}_{{\rm C}_2}^H\mathbf{A}_{{\rm R},2}\bm{\Gamma}_{1,2}\in\mathbb{C}^{N_{C_2,Y}\times G}$, $\text{r-supp}(\cdot)$ and
$\text{c-supp}(\cdot)$ are row-sparse support and column-sparse support of a sparse matrix, respectively.
 To solve it, a D-M-LAOMP algorithm is proposed by jointly recovering the row-block-spare matrices.

In the second stage with estimated $\mathbf{Z}_{2,1}$ and $\mathbf{Z}_{1,2}^H$, the non-zero rows of them are abstracted to form $\widetilde{\mathbf{Z}}_{2,1}$ and $\widetilde{\mathbf{Z}}_{1,2}^H$. Let the matrices formed by non-zero rows and non-zero columns of $\bm{\Gamma}_{2,1}$ and $\bm{\Gamma}_{1,2}$ be $\widetilde{\bm{\Gamma}}_{2,1}$ and $\widetilde{\bm{\Gamma}}_{1,2}$, respectively.
Then, we have $\widetilde{\mathbf{Z}}_{2,1}\approx\widetilde{\bm{\Gamma}}_{2,1}\mathbf{A}_{{\rm T},2}^H\mathbf{X}_{{\rm C}_2}$ and $\widetilde{\mathbf{Z}}_{1,2}\approx\widetilde{\mathbf{W}}^H_{{\rm C}_2}\mathbf{A}_{{\rm A},2}\widetilde{\bm{\Gamma}}_{1,2}$. This indicates how to recover the non-zero columns of $\bm{\Gamma}_{2,1}/\bm{\Gamma}_{1,2}^H$. Similar to problem (\ref{dmcs1}), the following D-M-CS problem is developed for the recovery of $\widetilde{\bm{\Gamma}}_{2,1}$ and $\widetilde{\bm{\Gamma}}_{1,2}$:
\begin{equation}\label{dmcs2}
	\setlength{\abovedisplayskip}{3.2pt}
	\begin{aligned}
		& \ \ \ 	\underset{\widetilde{\bm{\Gamma}}_{2,1},\widetilde{\bm{\Gamma}}_{1,2}}{\rm arg \ min} \ \left\Vert \widetilde{\bm{\Gamma}}_{2,1}\right\Vert_0+\left\Vert\widetilde{\bm{\Gamma}}_{1,2}\right\Vert_0\\
		& {\rm s.t.}\ \left\Vert\widetilde{\mathbf{Z}}_{2,1}^H-\widetilde{\mathbf{X}}^H_{{\rm C}_2}\mathbf{A}_{{\rm T},2}\widetilde{\bm{\Gamma}}_{2,1}^H\right\Vert_F^2\leq \epsilon,
		\\
		&\ \ \ \ \ \left\Vert\widetilde{\mathbf{Z}}_{1,2}-\widetilde{\mathbf{W}}^H_{{\rm C}_2}\mathbf{A}_{{\rm R},2}\widetilde{\bm{\Gamma}}_{1,2}\right\Vert_F^2\leq \epsilon,\\
		&\ \ \ \ \
		\text{ c-supp}( \widetilde{\bm{\Gamma}}_{2,1})=\text{r-supp}(\widetilde{\bm{\Gamma}}_{1,2}).
	\end{aligned}
\setlength{\belowdisplayskip}{3.2pt}
\end{equation}
The above problem can be effectively solved by D-M-LAOMP, which is easy to derive from our proposed D-LAOMP. 
According to the two-stage solutions, the BS's and UE's angles can be obtained by estimated $\text{r-supp}(\mathbf{Z}_{2,1})$ and $\text{c-supp}(\mathbf{\widetilde{\Gamma}}_{2,1})$.
\subsubsection{RIS's AoA/AoD Recovery (Angular-Domain Cascaded Channel Estimation)} With the estimated BS's and UE's angle directions, the BS and UE transmit/receive pilot signals along these angle directions. At the same time, the RIS adjusts its phase matrix constantly to estimate the RIS's channel parameters. For concreteness, let $\left\{\widehat{\bm{\Pi}}_{{\rm R},i},\widehat{\bm{\Pi}}_{{\rm T},i}|i=1,2\right\}$ be the estimated
angle parameters of the direct channel. Assuming that they are estimated approximately accurately, i.e., $\widehat{\bm{\Pi}}_{{\rm R},i}^H\bm{\Pi}_{{\rm R},i}\approx \mathbf{I}_{B_i}$ and $\widehat{\bm{\Pi}}_{{\rm T},i}^H\bm{\Pi}_{{\rm T},i}\approx \mathbf{I}_{B_i}$. In each sub-frame for RIS phase matrix adjustment, TX $i$ transmits pilot signals along $\bm{\Pi}_{{\rm T},i}$ and RX $j$ receives pilot signals with $\bm{\Pi}_{{\rm R},j}$. 
Similar to previous signal models, the collected signal at RX $j$ at the $q$-th sub-frame is 
\begin{equation}
	\setlength{\abovedisplayskip}{3.2pt}
	\begin{aligned}
\widehat{\mathbf{Y}}_{j,q}=&\bm{\Pi}_{{\rm R},j}^H\mathbf{H}_{{\rm R},j}\mathbf{V}_q\mathbf{H}_{{\rm T},i}\bm{\Pi}_{{\rm T},j}+\widetilde{\mathbf{N}}_{{\rm C}_j} \\ 
\approx&{\rm diag}(\bm{\chi}_j)\bm{\Pi}_{{\rm L},j}^H\mathbf{V}_q\bm{\Pi}_{{\rm L},i}{\rm diag}(\bm{\beta}_i)+\widetilde{\mathbf{N}}_{{\rm C}_j},
	\end{aligned}
\setlength{\belowdisplayskip}{3.2pt}
\end{equation}
where $\mathbf{V}_q$ is the RIS phase matrix in sub-frame $q$, $\widetilde{\mathbf{N}}_{{\rm C}_j}\in\mathbb{C}^{B_j\times B_i}$ is the noise matrix. For clarity, the noise term is ignored. The $(b_j,b_i)$-th element in $\widehat{\mathbf{Y}}_{j,q}$ is given by
\begin{equation}\label{YJQ}
	\begin{aligned}
\left[\widehat{\mathbf{Y}}_{j,q}\right]_{b_j,b_i} &\approx
[\bm{\chi}_j]_{b_j}\mathbf{v}_q\left(\mathbf{a}_l\left(\varrho_{j,b_j}^{\rm ele},\varrho_{j,b_j}^{\rm azi}\right)\odot\mathbf{a}_l^*\left(\varrho_{i,b_i}^{\rm ele},\varrho_{i,b_i}^{\rm azi}\right)\right)\left[\bm{\beta}_i\right]_{b_i}\\
&=[\bm{\chi}_j]_{b_j}\mathbf{v}_q\mathbf{c}_l\left(\varrho_{j,b_j}^{\rm ele},\varrho_{j,b_j}^{\rm azi},\varrho_{i,b_i}^{\rm ele},\varrho_{i,b_i}^{\rm azi}\right)\left[\bm{\beta}_i\right]_{b_i},
	\end{aligned}
\end{equation}
where $\mathbf{c}_l$ is a cascaded AR formed by the Hardmard product of two PARs. By the way, we can define the angular-domain cascaded channel $\bm{\Xi}_{i,j}\in\mathbb{C}^{L\times B_jB_i}$, which will be used in a later section for passive beamforming,
 with the $((b_i-1)B_j+b_j)$-column being
\begin{equation}\label{Hcas}
[\bm{\Xi}_{i,j}]_{:,(b_i-1)B_j+b_j}\triangleq \mathbf{c}_l\left(\varrho_{j,b_j}^{\rm ele},\varrho_{j,b_j}^{\rm azi},\varrho_{i,b_i}^{\rm ele},\varrho_{i,b_i}^{\rm azi}\right)\left[\bm{\beta}_i\right]_{b_i}[\bm{\chi}_j]_{b_j}.
\end{equation}

Then, the $((b_i-1)B_j+b_j)$-the element of ${\rm vec}\left(\widehat{\mathbf{Y}}_{j,q}\right)$ is written as
\begin{equation}
	\begin{aligned}
\left[{\rm vec}\left(\widehat{\mathbf{Y}}_{j,q}\right)\right]_{(b_i-1)B_j+b_j}=\mathbf{v}_q\mathbf{c}_l\left(\varrho_{j,b_j}^{\rm ele},\varrho_{j,b_j}^{\rm azi},\varrho_{i,b_i}^{\rm ele},\varrho_{i,b_i}^{\rm azi}\right)\left[\bm{\tau}_{i,j}\right]_{(b_i-1)B_j+b_j},
	\end{aligned}
\end{equation}
where $\bm{\tau}_{i,j}=\bm{\beta}_i\otimes\bm{\chi}_j\in\mathbb{C}^{1\times B_iB_j}$ is the cascaded path gain. After $Q$ sub-frames, the collected signal at RX $j$ is denoted by $\mathbf{\ddot{Y}}_j\in\mathbb{C}^{B_iB_j\times Q}$. The $((b_i-1)B_j+b_j)$-th column of its transpose $\mathbf{\ddot{Y}}_j^T$ is given by
\begin{equation}\label{DDY}
	\begin{aligned}
\left[\mathbf{\ddot{Y}}_j^T\right]_{:,(b_i-1)B_j+b_j} =\overline{\mathbf{V}}\mathbf{c}_l\left(\varrho_{j,b_j}^{\rm ele},\varrho_{j,b_j}^{\rm azi},\varrho_{i,b_i}^{\rm ele},\varrho_{i,b_i}^{\rm azi}\right)\left[\bm{\tau}_{i,j}\right]_{(b_i-1)B_j+b_j},
	\end{aligned}
\end{equation}
where $\overline{\mathbf{V}}=\left[\mathbf{v}_1^T,\mathbf{v}_2^T,\cdots,\mathbf{v}_Q^T\right]^T\in\mathbb{C}^{Q\times L}$ is the collected RIS phase matrix for estimating the RIS's channel parameters. This equation shows that each column of $\mathbf{\ddot{Y}}_j^T$ is composed of one AR and one path gain. Thus,
we can recover $\mathbf{c}_l\left(\varrho_{j,b_j}^{\rm ele},\varrho_{j,b_j}^{\rm azi},\varrho_{i,b_i}^{\rm ele},\varrho_{i,b_i}^{\rm azi}\right)$ and $\left[\bm{\tau}_{i,j}\right]_{(b_i-1)B_j+b_j}$ from the $((b_i-1)B_j+b_j)$-th column of $\mathbf{\ddot{Y}}_j^T$ by designing a dictionary which samples the RIS's beamspace. Let $\mathbf{C}_{\rm L}\in\mathbb{C}^{L \times G_{\rm L}}$ be a dictionary with $G_{\rm L}$ atoms, characterizing the RIS's beamspace from different angle directions, Eqn. (\ref{DDY}) can be approximated as
\begin{equation}
\left[\mathbf{\ddot{Y}}_j^T\right]_{:,(b_i-1)B_j+b_j} \approx \overline{\mathbf{V}}\mathbf{C}_{\rm L}[\mathbf{t}_{i,j}]_{(b_i-1)B_j+b_j},
\end{equation}
where $[\mathbf{t}_{i,j}]_{(b_i-1)B_j+b_j}\in\mathbb{C}^{G_{\rm L}\times 1}$ is a $1$-sparse vector. According to this, $[\mathbf{t}_{i,j}]_{(b_i-1)B_j+b_j}\in\mathbb{C}^{G_{\rm L}\times 1}$ can be recovered by 
the standard CS framework, i.e., for $j=1,2 \ (i\in\{1,2\},i\neq j)$,
\begin{equation}
	\begin{aligned}
 \underset{[\mathbf{t}_{i,j}]_{(b_i-1)B_j+b_j}}{\rm arg \ min}\left\Vert [\mathbf{t}_{i,j}]_{(b_i-1)B_j+b_j}\right\Vert_0 , \ {\rm s.t.} \
\left\Vert\left[\mathbf{\ddot{Y}}_j^T\right]_{:,(b_i-1)B_j+b_j}-\overline{\mathbf{V}}\mathbf{C}_{\rm L}[\mathbf{t}_{i,j}]_{(b_i-1)B_j+b_j}\right\Vert_2^2\leq \epsilon.
	\end{aligned}
\end{equation}
By recovering $\widehat{\mathbf{t}}_{i,j}$, the angular-domain cascaded channel $\widehat{\bm{\Xi}}_{i,j}$ can be reconstructed.
The above problem can be directly solved by LAOMP. For this, the dictionary $\mathbf{C}_{\rm L}$ should be designed. Since the AR $\mathbf{c}_l$ of Eqn. (\ref{DDY}) consists of four parameters, sampling them to characterize the beamspace will produce an immense dictionary. Note that $\mathbf{c}_l$ is the Hardmard product of two ARs. Hence $\mathbf{c}_l\left(\varrho_{j,b_j}^{\rm ele},\varrho_{j,b_j}^{\rm azi},\varrho_{i,b_i}^{\rm ele},\varrho_{i,b_i}^{\rm azi}\right)$ has a form of
\begin{equation}\label{al}
	\begin{aligned}
\mathbf{c}_l\left(\varrho_{j,b_j}^{\rm ele},\varrho_{j,b_j}^{\rm azi},\varrho_{i,b_i}^{\rm ele},\varrho_{i,b_i}^{\rm azi}\right)=\mathbf{a}_l\left(\varrho_{j,b_j}^{\rm ele}-\varrho_{i,b_i}^{\rm ele},\varrho_{j,b_j}^{\rm azi}-\varrho_{i,b_i}^{\rm azi}\right)=\mathbf{a}_l 
\left(\rho^{\rm ele}_{i,j,b_i,b_j},\rho^{\rm azi}_{i,j,b_i,b_j}\right),
	\end{aligned}
\end{equation}
where $\rho^{\rm ele}_{i,j,b_i,b_j}\triangleq\varrho_{j,b_j}^{\rm ele}-\varrho_{i,b_i}^{\rm ele}$ and $\rho^{\rm azi}_{i,j,b_i,b_j}\triangleq\varrho_{j,b_j}^{\rm azi}-\varrho_{i,b_i}^{\rm azi}$ serve as the cascaded elevation and azimuth angles of the RIS, respectively. 
Note that cascaded angles belong to $[-2,2]$ since each angle in $\{\varrho_{j,b_j}^{\rm ele},\varrho_{j,b_j}^{\rm azi},\varrho_{i,b_i}^{\rm ele},\varrho_{i,b_i}^{\rm azi}\}$ belongs to $[-1,1]$. However, we just need to sample the cascaded angle in $[-1,1]$ due to the exponential terms $e^{j\pi\frac{2d}{\lambda} (\varrho_{j,b_j}^{\rm ele}-\varrho_{i,b_i}^{\rm ele})}$ and $e^{j\pi\frac{2d}{\lambda} (\varrho_{j,b_j}^{\rm azi}-\varrho_{i,b_i}^{\rm azi})}$, where $d=\frac{\lambda}{2}$ is assumed. Therefore, the dictionary $\mathbf{C}_{\rm L}$ can be represented by a over-complete DFT dictionary as shown in Section \ref{VCR}.
\vspace{-0.4cm}
\subsection{Off-Grid Refinement}
Due to the basis mismatch problem \cite{mismatch} inherent in AoA/AoD estimation, the off-grid or super-resolution optimization after CS recovery should be considered for better estimation performance. Indeed, some recovery algorithms themselves can resist the basis mismatch impact, e.g., atomic norm minimization \cite{ANM}, but with a high computational complexity. Thus, the preferred fast method is off-grid refinement that jointly optimizes the selected atoms and non-zero coefficients. Here, the off-grid method in our previous works \cite{OG1,OG2} is used for the proposed CS frameworks. The key idea is to utilize the perturbation technique to minimize the error between the estimated angle and the true angle. Although it is not direct to design different off-grid methods for variant LAOMP algorithms, e.g., off-grid D-M-LAOMP for Eqn. (\ref{dmcs1}) and (\ref{dmcs2}), the MMV-version off-grid method in \cite{OG2} can be referred to for further extension.
\vspace{-0.2cm}
\section{Passive and Hybrid Beamfoming Optimization}\label{Sec5}
\vspace{-0.2cm}
 Since the accurate separate channels $\{\mathbf{H}_{{\rm R},j},\mathbf{H}_{{\rm T},i}\}$ are hard to acquire  due to 1) the ambiguity problem \cite{AMP}, and 2) the hybrid beamforming architecture, this section conducts passive beamforming optimization based on the angular-domain cascaded channel of Eqn. (\ref{Hcas}).
With optimized passive beamforming, a H-WMMSE-SIC method that iteratively  optimizes hybrid beamforming is proposed.
\vspace{-0.5cm}
\subsection{Angular-Domain-Based Passive Beamforming }

 Let  $\bm{\Sigma}_j\triangleq\sigma_{n,j}^2\mathbf{W}^H_{j}\mathbf{W}_{ j}$ and $\sigma_{p,i}^2$ be the processed noise covariance matrix at RX $j$ and the  transmit power at TX $i$, respectively. By optimizing the cascaded channel's SE, the following optimization problem is formulated:
\begin{equation}\label{SPC}
	\setlength{\abovedisplayskip}{3.2pt}
	\begin{aligned}
& \underset{\mathbf{V}}{\rm arg \ max} \ \sum_{j=1}^{2}	{\rm log}_2\left({\rm det}\left(\mathbf{I}_{N_{st,i}}+\bm{\Sigma}^{-1}_j\mathbf{W}^H_{j}\mathbf{H}_{{\rm R}, j}\mathbf{V}\mathbf{H}_{{\rm T},i} \mathbf{F}_{i} 
\mathbf{F}^H_{i}\mathbf{H}_{{\rm T}, i}^H\mathbf{V}^H\mathbf{H}_{{\rm R},j}^H\mathbf{W}_{j}\right)\right)
\\&  \ \ \ \   {\rm s.t.} \ 
\vert[\mathbf{V}]_{l_1,l_2}\vert=\begin{cases}
1, l_1=l_2\\
0,l_1\neq l_2
\end{cases} \left(\forall l_1,l_2=\{1,\cdots, L\}\right), \ {\rm Tr}(\mathbf{F}_i\mathbf{F}_i^H)\leq \sigma_{p,i}^2.
	\end{aligned}
\setlength{\belowdisplayskip}{3.2pt}
\end{equation}
Suppose that $\{\mathbf{F}_{i},\mathbf{W}_{j}\}$ are fixed as the optimal fully-digital precoder/combiner such that $\mathbf{F}_i^H\mathbf{F}_i=\frac{\sigma_{p,i}^2}{N_{st,i}}\mathbf{I}_{N_{st,i}}$ and $\mathbf{W}_j^H\mathbf{W}_j=\mathbf{I}_{N_{st,i}}$. Therefore, maximizing the above objective function is equivalent to maximizing 
$
{\rm SE}_{{\rm cas},j}\triangleq{\rm log}_2\left({\rm det}\left(\mathbf{I}_{N_{st,i}}+\frac{\sigma_{p,i}^2}{N_{st,i}}\bm{\Sigma}_j^{-1}\mathbf{H}_{{\rm R},j}\mathbf{V}\mathbf{H}_{{\rm T},i}\mathbf{H}_{{\rm T},i}^H\mathbf{V}^H
\mathbf{H}_{{\rm R},j}^H\right)\right)
$.
However, it is still difficult to handle due to the unknown channel matrices $\{\mathbf{H}_{{\rm R},j},\mathbf{H}_{{\rm T},i}\}$, the determinant
operation, as well as the non-convex
constraint of the RIS phase matrix. 
To address the optimization problem of Eqn. (\ref{SPC}) without known $\{\mathbf{H}_{{\rm R},j},\mathbf{H}_{{\rm T},i}\}$, as Eqn. (\ref{YCJ}), we re-write ${\rm SE}_{{\rm cas},j}={\rm log}_2\left({\rm det}\left(\mathbf{I}_{N_{st,i}}+\frac{\sigma_{p,i}^2}{N_{st,i}\sigma_n^2}\bm{\Pi}_{{\rm R},j} \bm{\Lambda}_{i,j}\bm{\Pi}_{{\rm T},i}^H\bm{\Pi}_{{\rm T},i}\bm{\Lambda}_{i,j}^H\bm{\Pi}_{{\rm R},j}^H\right)\right)$. 
 Assuming that the number of data streams equals to the rank of the cascaded channel.
 The upper bound of ${\rm SE}_{{\rm cas},j}$ is
\begin{equation}
	\setlength{\abovedisplayskip}{3.2pt}
	\begin{aligned}
{\rm SE}_{{\rm cas},j}&= {\rm log}_2\left(\prod_{k=1}^{{N_{st,i}}}(1+\frac{\sigma_{p,i}^2}{N_{st,i}\sigma_{n,j}^2}\lambda_k^2)\right)\\ 
 &\overset{(e)}{\leq}{N_{st,i}}{\rm log}_2\left(1+\frac{\sigma_{p,i}^2{\rm Tr}\left(\bm{\Pi}_{{\rm R},j} \bm{\Lambda}_{i,j}\bm{\Pi}_{{\rm T},i}^H\bm{\Pi}_{{\rm T},i}\bm{\Lambda}_{i,j}^H\bm{\Pi}_{{\rm R},j}^H \right)}{N^2_{st,i}\sigma_{n,j}^2}\right)\\
 &\overset{(f)}{\leq}{N_{st,i}}{\rm log}_2\left(1+\bm{\digamma}_{i,j}{\rm Tr}\left( \bm{\Lambda}_{i,j}\bm{\Lambda}_{i,j}^H\right) \right),
	\end{aligned}
\setlength{\belowdisplayskip}{3.2pt}
\end{equation}
where $\{\lambda_k\}_{k=1}^{N_{st,i}}$ are singular values of $\bm{\Pi}_{{\rm R},j} \bm{\Lambda}_{i,j}\bm{\Pi}_{{\rm T},i}^H$,  $\bm{\digamma}_{i,j}\triangleq\frac{\sigma_{p,i}^2}{N^2_{st,i}\sigma_{n,j}^2}{\rm Tr}(\bm{\Pi}_{{\rm T},i}^H\bm{\Pi}_{{\rm T},i}){\rm Tr}(\bm{\Pi}_{{\rm R},j}^H\bm{\Pi}_{{\rm R},j})$, $(e)$ holds owing to the Jensen's inequality, and $(f)$ holds due to ${\rm Tr}(\mathbf{AB})\leq {\rm Tr}(\mathbf{A}){\rm Tr}(\mathbf{B})$ with arbitrary matrics $\mathbf{A}\succeq 0$ and $\mathbf{B}\succeq 0$.
Therefore, maximizing Eqn. (\ref{SPC}) is equivalent to maximizing ${\rm Tr}\left( \bm{\Lambda}_{1,2}\bm{\Lambda}_{1,2}^H\right)+{\rm Tr}\left( \bm{\Lambda}_{2,1}\bm{\Lambda}_{2,1}^H\right)$ w.r.t. $\mathbf{V}\triangleq{\rm diag}(\mathbf{v})$. By a simple transformation of ${\rm Tr}\left( \bm{\Lambda}_{i,j}\bm{\Lambda}_{i,j}^H\right)=\left\Vert \bm{\Lambda}_{i,j}\right\Vert_F^2$ and combining Eqn. (\ref{LAM}), (\ref{YJQ}), and (\ref{Hcas}), we have
$
{\rm Tr}\left( \bm{\Lambda}_{i,j}\bm{\Lambda}_{i,j}^H\right)=\left\Vert \mathbf{v}\widehat{\bm{\Xi}}_{i,j} \right\Vert_2^2
$,
which motivates us to propose an angular-domain-based optimization problem w.r.t. $\mathbf{v}$, which is equivalent  to problem (\ref{SPC}), given by
\begin{equation}
	\begin{aligned}
 \underset{\mathbf{v}}{\rm arg \ max } \ \mathbf{v}\left(\widehat{\bm{\Xi}}_{1,2}\widehat{\bm{\Xi}}_{1,2}^H+\widehat{\bm{\Xi}}_{2,1}\widehat{\bm{\Xi}}_{2,1}^H\right)\mathbf{v}^H , \   {\rm s.t.} \ \vert[\mathbf{v}]_{k_1}\vert=1, \forall k_1=1,\cdots,L.
	\end{aligned}
\end{equation}
Since the above problem is non-convex regardless of the constraint of $\mathbf{v}$,
singular value decomposition (SVD) is used to find the largest singular vector for a sub-optimal solution. Using SVD on the Hermitian positive semi-definite matrix $\widehat{\bm{\Xi}}_{1,2}\widehat{\bm{\Xi}}_{1,2}^H+\widehat{\bm{\Xi}}_{2,1}\widehat{\bm{\Xi}}_{2,1}^H=\mathbf{\ddot{V}}\bm{\Sigma}_{\mathbf{v}}\mathbf{\ddot{V}}^H$, where $\bm{\Sigma}_{\mathbf{v}}$ is a diagonal matrix with singular values in descending order.
With the modulo 1 constraint, $\mathbf{v}^\star$ can be derived by $\mathbf{v}^\star=e^{j(-{\rm angle}(\mathbf{\ddot{V}}_{:,1}))}$, where ${\rm angle}(\cdot)$ denotes the phase abstraction operator. Finally, we can obtain $\widehat{\bm{\Lambda}}_{i,j}$ with optimized $\mathbf{v}^\star$ and estimated $\widehat{\bm{\Xi}}_{i,j}$.


\vspace{-0.5cm}
\subsection{H-WMMSE-SIC}

By denoting ${\mathbf{H}}_{{\rm C},i}\triangleq{\bm{\Pi}}_{{\rm R},j} {\bm{\Lambda}}_{i,j}{\bm{\Pi}}_{{\rm T},i}^H$,
the SE at RX $j$ is defined as 
\begin{equation}
	\begin{aligned}
{\rm SE}_{j}\triangleq{\rm log}_2\left({\rm det}\left(\mathbf{I}_{N_{st,i}}+\widetilde{\bm{\Sigma}}^{-1}_j\mathbf{W}^H_{j}\widehat{\mathbf{H}}_{{\rm DC},i}\mathbf{F}_{i}\mathbf{F}_{i}^H\widehat{\mathbf{H}}_{{\rm DC},i}^H\mathbf{W}_{j} \right)\right),
	\end{aligned}
\end{equation}
with ${\mathbf{H}}_{{\rm DC},i}={\mathbf{H}}_{{\rm D},i}+{\mathbf{H}}_{{\rm C},i}$ and $\widetilde{\bm{\Sigma}}_j=\sigma_{n,j}^2\mathbf{W}_j^H\mathbf{W}_j+\mathbf{W}_j^H \mathbf{H}_{{\rm S},j}\mathbf{F}_j\mathbf{F}_j^H\mathbf{H}_{{\rm S},j}^H\mathbf{W}_j$. 
Denoting the hybrid beamfoming set as
$\mathcal{H}=\{\mathbf{F}_{{\rm RF},i},\mathbf{F}_{{\rm BB},i},\mathbf{W}_{{\rm RF},i},\mathbf{W}_{{\rm BB},i}|i=1,2\}
$,
the optimization problem is written as 
\begin{equation}\label{SEJ}
	\setlength{\abovedisplayskip}{3.2pt}
\begin{aligned}
&\underset{\mathcal{H}}{\rm arg \ max}\ {\rm SE}_{1}+{\rm SE}_{2} \\
 {\rm s.t.} \ 
&\vert[\mathbf{F}_{{\rm RF},i}]_{k_1,k_2}\vert  = \vert[\mathbf{W}_{{\rm RF},i}]_{k_1,k_2}\vert=1, \ \forall k_1,k_2,\\&
\left\Vert\mathbf{F}_{{\rm RF},i}\mathbf{F}_{{\rm BB},i}\right\Vert_F^2\leq \sigma_{p,i}^2, \forall i=1,2,
\end{aligned}
\setlength{\belowdisplayskip}{3.2pt}
\end{equation}
The above problem is hard to solve due to the non-convex variables in $\mathcal{H}$ and the multiplication of those non-convex variables and the SI term. Inspired by the WMMSE minimization method used for broadcast channel precoder/combiner \cite{WMMSE} with a good performance, we employ it for SIC.
 Based on the relation between the SE
maximization and WMMSE minimization problems, problem (\ref{SEJ}) is transformed into a more
tractable equivalent form:
\begin{equation}\label{QE}
	\setlength{\abovedisplayskip}{3.2pt}
	\begin{aligned}
\underset{\mathcal{M}}{\rm arg \ min} &\ \sum_{j=1}^{2} {\rm Tr}\left(\mathbf{Q}_j\mathbf{E}_j\right)-{\rm log}_2({\rm det}(\mathbf{Q}_j)) \\
 {\rm s.t.} \ 
&\vert[\mathbf{F}_{{\rm RF},i}]_{k_1,k_2}\vert  = \vert[\mathbf{W}_{{\rm RF},i}]_{k_1,k_2}\vert=1, \ \forall k_1,k_2,\\&
\left\Vert\mathbf{F}_{{\rm RF},i}\mathbf{F}_{{\rm BB},i}\right\Vert_F^2\leq \sigma_{p,i}^2, \forall i=1,2,
	\end{aligned}
\setlength{\belowdisplayskip}{3.2pt}
\end{equation}
where $\mathcal{M}=\{\mathcal{H},\mathbf{Q}_1,\mathbf{Q}_2\}$, and $\mathbf{Q}_j\succeq 0$ is a weighted matrix for RX $j$. The MSE covariance matrix is defined as
\begin{equation}\label{EJ}
	\begin{aligned}
\mathbf{E}_j&\triangleq\mathbb{E}\left[ (\mathbf{s}_j-\widehat{\mathbf{s}}_j)(\mathbf{s}_j-\widehat{\mathbf{s}}_j)^H\right]\\
&=\left(\mathbf{I}_{N_{st,i}}-\mathbf{W}^H_j{\mathbf{H}}_{{\rm DC},i}\mathbf{F}_i\right)\left(\mathbf{I}_{N_{st,i}}-\mathbf{W}^H_j{\mathbf{H}}_{{\rm DC},i}\mathbf{F}_i\right)^H +\mathbf{W}_j^H {\mathbf{H}}_{{\rm S},j}\mathbf{F}_j\mathbf{F}_j^H{\mathbf{H}}_{{\rm S},j}^H\mathbf{W}_j+\sigma_{n,j}^2\mathbf{W}_j^H\mathbf{W}_j,
	\end{aligned}
\end{equation}
where $\widehat{\mathbf{s}}_j$ is the combined received signal at RX $j$.
With the above equivalence, all variables can be iteratively solved with the fully-digital array architecture, i.e., the fully-digital WMMSE-SIC method. However, the non-convex variable set $\mathcal{H}$ is still hard to solve. One straightforward method is to obtain a fully-digital solution, followed by a decoupling of analog and digital beamforming. This way will incur large performance loss due to the strong SI. To meet this drawback, we propose a H-WMMSE-SIC method which optimizes $\mathcal{H}$ in each iteration of the WMMSE procedure.

\subsubsection{Optimization of $\{\mathbf{W}_{{\rm RF},j},\mathbf{W}_{{\rm BB},j}\}_{j=1}^2$ }\label{OW}
By fixing other variables except $\{\mathbf{W}_{{\rm RF},j},\mathbf{W}_{{\rm BB},j}\}_{j=1}^2$, the optimization problem of (\ref{QE}) is equivalent to minimizing $\mathbb{E}[\Vert\mathbf{s}-\widehat{\mathbf{s}}\Vert_2^2]$.
 For $j=1,2 \ (i\in\{1,2\},i\neq j)$, we have
 \begin{equation}\label{ESW}
\begin{aligned}
\underset{\mathbf{W}_{{\rm RF},j},\mathbf{W}_{{\rm BB},j} }{\rm arg \ min}\ \mathbb{E}\left[\Vert\mathbf{s}_j-\widehat{\mathbf{s}}_j\Vert_2^2\right], \ {\rm s.t.} \ \vert[\mathbf{W}_{{\rm RF},i}]_{k_1,k_2}\vert=1, \ \forall k_1,k_2.
\end{aligned}
 \end{equation}
The above objective function can be further written as
\begin{equation}\label{ES}
	\begin{aligned}
\mathbb{E}\left[\Vert\mathbf{s}_j-\widehat{\mathbf{s}}_j\Vert_2^2\right]=&{\rm Tr}\left(\mathbf{W}_{{\rm BB},j}^H\mathbf{W}_{{\rm RF},j}^H\mathbf{U}_j\mathbf{W}_{{\rm RF},j}\mathbf{W}_{{\rm BB},j}\right) \\ &
 -2\Re\left\{{\rm Tr}\left(\mathbf{W}_{{\rm BB},j}^H\mathbf{W}_{{\rm RF},j}^H{\mathbf{H}}_{{\rm DC},i}\mathbf{F}_{{\rm RF},i}\mathbf{F}_{{\rm BB},i}\right)\right\}+1 ,
	\end{aligned}
\end{equation}
 where $\mathbf{U}_j={\mathbf{H}}_{{\rm DC},i}\mathbf{F}_{{\rm RF},i}\mathbf{F}_{{\rm BB},i}\mathbf{F}_{{\rm BB},i}^H\mathbf{F}_{{\rm RF},i}^H{\mathbf{H}}_{{\rm DC},i}^H+{\mathbf{H}}_{{\rm S},j}\mathbf{F}_{{\rm RF},j}\mathbf{F}_{{\rm BB},j}\mathbf{F}_{{\rm BB},j}^H\mathbf{F}_{{\rm RF},j}^H{\mathbf{H}}_{{\rm S},j}^H+\sigma_{n,j}^2\mathbf{I}_{N_{j,r}}$. 
If there are no constraints on $\mathbf{W}_{{\rm RF},j}\mathbf{W}_{{\rm BB},j}$ (i.e., fully-digital WMMSE-SIC is considered), the optimal soluition can be derived by
the first-order optimality condition. This indicates that the optimal $\mathbf{W}_{{\rm BB},j}^\star$ can be directly solved by fixing $\mathbf{W}_{{\rm RF},j}$. According to
the first-order optimality condition, we have 
\begin{equation}\label{WBBX}
\mathbf{W}_{{\rm BB},j}^\star=\left(\mathbf{W}_{{\rm RF},j}^H\mathbf{U}_j\mathbf{W}_{{\rm RF},j}\right)^{-1}\mathbf{W}_{{\rm RF},j}{\mathbf{H}}_{{\rm DC},i}\mathbf{F}_{{\rm RF},i}\mathbf{F}_{{\rm BB},i}.
\end{equation}
With this, we can minimize Eqn. (\ref{ES}) w.r.t. $\mathbf{W}_{{\rm RF},j}$ subject to ${\rm s.t.} \ \vert[\mathbf{W}_{{\rm RF},i}]_{k_1,k_2}\vert=1$. Taking into account the modulus constraint of each entry in $\mathbf{W}_{{\rm RF},j}$, the following proposition aims at finding each entry's solution while fixing other entries.

\emph{Proposition 2: Considering a  problem given by
 \begin{equation}
	\begin{aligned}
		&\underset{\mathbf{R} }{\rm arg \ min}\ {\rm Tr}\left(
		\mathbf{B}^H\mathbf{R}^H\mathbf{U}\mathbf{R}\mathbf{B}\right)-2\Re\left\{{\rm Tr}\left\{\mathbf{B}^H\mathbf{R}^H\mathbf{F}\right\}
		\right\}
		 \\
		&  \ \ \ \ \ \ \ \ \ \ \ \ \ \ {\rm s.t.} \ \vert[\mathbf{R}]_{k_1,k_2}\vert=1, \ \forall k_1,k_2.
	\end{aligned}
\end{equation}
Its solution can be derived by the CD procedure, which optimizes each entry of $\mathbf{R}$ while fixing other entries, i.e., $[\mathbf{R}]_{k_1,k_2}=\frac{\nu_{k_1,k_2}}{\vert\nu_{k_1,k_2}\vert}$ with $\nu_{k_1,k_2}=\sum_{k_3\neq k1}\sum_{k_4\neq k2}[\mathbf{U}]_{k_1,k_3}[\mathbf{R}]_{k_3,k_4}[\mathbf{BB}^H]_{k_4,k_2}-[\mathbf{FB}^H]_{k_1,k_2}$.
}

\emph{Proof:} See Appendix \ref{appendixB}.

According to the above proposition, $\mathbf{W}_{{\rm RF},j}^\star$ can be derived by solving Eqn. (\ref{ESW}) with fixed $\mathbf{W}_{{\rm BB},j}^\star$. 
 Next, we update the weighted matrix for WMMSE.

\subsubsection{Optimization of $\{\mathbf{Q}_j\}_{j=1}^2$}
With the fixed hybrid beamforming set $\mathcal{H}$, problem (\ref{QE}) is convex w.r.t. $\mathbf{Q}_j$. Hence
the optimal weighted matrix can be derived as $\mathbf{Q}_j^\star=\mathbf{E}_j^{-1}$  by the first-order optimality condition.

\subsubsection{Optimization of $\{\mathbf{F}_{{\rm RF},i},\mathbf{F}_{{\rm BB},i}\}_{i=1}^2$}\label{OF}
For convenience of understanding, we first derive the fully-digital solution for $\mathbf{F}_i$ when fixing the weighted matrices and the hybrid precoder. The following problem is formulated:
\begin{equation}
	\setlength{\abovedisplayskip}{3.2pt}
\begin{aligned}
& \underset{\mathbf{F}_1,\mathbf{F}_2}{\rm arg \ min} \ {\rm Tr}\left(\mathbf{Q}_1\mathbf{E}_1\right)+{\rm Tr}\left(\mathbf{Q}_2\mathbf{E}_2\right) \\
&  \ \ \ {\rm s.t.} \ \vert [\mathbf{F}_{{\rm RF},j}]_{k_1,k_2}\vert =1,  \ \forall k_1,k_2, \\
&\ \ \ \  \left\Vert\mathbf{F}_{{\rm RF},i}\mathbf{F}_{{\rm BB},i}\right\Vert_F^2\leq \sigma_{p,i}^2,\forall i=1,2.
\end{aligned}
\setlength{\belowdisplayskip}{3.2pt}
\end{equation}
The above objective function is expanded as
\begin{equation}
\setlength{\abovedisplayskip}{3.2pt}
	\begin{aligned}
{\rm Tr}\left(\mathbf{Q}_1\mathbf{E}_1\right)+&{\rm Tr}\left(\mathbf{Q}_2\mathbf{E}_2\right)=
{\rm Tr}\left(\mathbf{F}_1^H\mathbf{T}_1\mathbf{F}_1+\mathbf{F}_2^H\mathbf{T}_2\mathbf{F}_2
\right) +{\rm Tr}\left(\mathbf{Q}_1\mathbf{W}_1^H\mathbf{W}_1+\mathbf{Q}_2\mathbf{W}_2^H\mathbf{W}_2\right)   \\ 
&\ \ \ \ \ -2\Re\left\{{\rm Tr}\left(\mathbf{F}_1^H{\mathbf{H}}_{{\rm DC},1}^H \mathbf{W}_2\mathbf{Q}_2+\mathbf{F}_2^H{\mathbf{H}}_{{\rm DC},2}^H \mathbf{W}_1\mathbf{Q}_1\right)\right\}  +{\rm Tr}\left(\mathbf{Q}_1+\mathbf{Q}_2\right),
	\end{aligned}
\setlength{\belowdisplayskip}{3.2pt}
\end{equation}
where $\mathbf{T}_i={\mathbf{H}}_{{\rm DC},i}^H\mathbf{W}_j
\mathbf{Q}_j\mathbf{W}_j^H{\mathbf{H}}_{{\rm DC},i} + {\mathbf{H}}_{{\rm S},i}^H\mathbf{W}_i\mathbf{Q}_i\mathbf{W}_i^H{\mathbf{H}}_{{\rm S},i}$ for $i,j=1,2, j\neq i$. 
According to this, the optimization for precoder can be decoupled across the two TXs. For $i=1,2, \ (j=1,2, j\neq i)$, we optimize the following objective function
\begin{equation}\label{FT}
	\setlength{\abovedisplayskip}{3.2pt}
{\rm Tr}\left(\mathbf{F}_i^H\mathbf{T}_i\mathbf{F}_i\right)-2\Re\left\{{\rm Tr}\left(\mathbf{F}_i^H{\mathbf{H}}_{{\rm DC},i}^H \mathbf{W}_j\mathbf{Q}_j\right)\right\},
\setlength{\belowdisplayskip}{3.2pt}
\end{equation}
subject to $\mathbf{F}_{{\rm RF},i}\in \mathbf{A}_{{\rm T},i}$.
If there are no constraints on $\mathbf{F}_i$, the above problem w.r.t. $\mathbf{F}_i$ is convex due to $\mathbf{T}_i\succeq 0$. Hence $\mathbf{F}_i^\star$ can be obtained by the Lagrangian duality, where $\mathbf{F}_i^\star=\left(\mathbf{T}_i+\mu_i\mathbf{I}_{N_{i,t}}\right)^{-1}{\mathbf{H}}_{{\rm DC},i}^H \mathbf{W}_j\mathbf{Q}_j$, where $\mu_i$ is the factor that controls the power constraint $\left\Vert\mathbf{F}_i^\star\right\Vert_F^2\leq \sigma_{p,i}^2$. 
 Since $\left\Vert\mathbf{F}_i^\star\right\Vert_F^2$ is a monotonically decreasing function of $\mu_i$. Thus, the bisection algorithm can be used for solving a proper $\mu_i$.
 However, the hybrid precoder is still hard to get based on minimizing Eqn. (\ref{FT}). 
For this, $\mathbf{F}_{{\rm BB},i}$ and $\mathbf{F}_{{\rm RF},i}$ are iteratively optimized. By fixing $\mathbf{F}_{{\rm RF},i}$, optimal $\mathbf{F}_{{\rm BB},i}^{\star}$ can be obtained as the derivation of $\mathbf{F}_i^\star$, i.e.,
\begin{equation}
	\setlength{\abovedisplayskip}{3.2pt}
\mathbf{F}_{{\rm BB},i}^{\star}=\left(\widetilde{\mathbf{T}}_i+\widetilde{\mu}_i\mathbf{F}_{{\rm RF},i}^H\mathbf{F}_{{\rm RF},i}\right)^{-1}\mathbf{F}_{{\rm RF},i}^H{\mathbf{H}}_{{\rm DC},i}^H \mathbf{W}_j\mathbf{Q}_j,
\setlength{\belowdisplayskip}{3.2pt}
\end{equation}
where $\widetilde{\mathbf{T}}_i=\mathbf{F}_{{\rm RF},i}^H{\mathbf{H}}_{{\rm DC},i}^H\mathbf{W}_j
\mathbf{Q}_j\mathbf{W}_j^H{\mathbf{H}}_{{\rm DC},i}\mathbf{F}_{{\rm RF},i} + \mathbf{F}_{{\rm RF},i}^H{\mathbf{H}}_{{\rm S},i}^H\mathbf{W}_i\mathbf{Q}_i\mathbf{W}_i^H{\mathbf{H}}_{{\rm S},i}\mathbf{F}_{{\rm RF},i}$, and $\widetilde{\mu}_i$ is the power controlling factor making $\left\Vert\mathbf{F}_{{\rm RF},i}\mathbf{F}_{{\rm BB},i}^\star\right\Vert_F^2\leq \sigma_{p,i}^2$. Subsequently, we can
 utilize proposition 2 to optimize $\mathbf{F}_{{\rm RF},i}$ by solving
 \begin{equation}
 	\setlength{\abovedisplayskip}{3.2pt}
 	\begin{aligned}
 	&	\underset{\mathbf{F}_{{\rm RF},i}}{\rm arg \ min} \
{\rm Tr}\left(\mathbf{F}_{{\rm BB},i}^{\star,H}\mathbf{F}_{{\rm RF},i}^{H}\mathbf{T}_i\mathbf{F}_{{\rm RF},i}\mathbf{F}_{{\rm BB},i}^\star\right) -2\Re\left\{{\rm Tr}\left(\mathbf{F}_{{\rm BB},i}^{\star,H}\mathbf{F}_{{\rm RF},i}^{H}{\mathbf{H}}_{{\rm DC},i}^H \mathbf{W}_j\mathbf{Q}_j\right)\right\}\\
& \ \ \ \ \ \ \ \  \ \ \ \  \ \ \ \ \ \ \ \ \ \ \  \ \ \ {\rm s.t.} \ \vert[\mathbf{F}_{{\rm RF},i}]_{k_1,k_2}\vert=1, \ \forall k_1,k_2.
 	\end{aligned}
 \setlength{\belowdisplayskip}{3.2pt}
 \end{equation}

To this point, the hybrid beamforming set $\mathcal{H}$ can be iteratively optimized according to Algorithm \ref{HWMMSE}. 
\begin{algorithm}[!t] 
	\caption{H-WMMSE-SIC } 
	\label{HWMMSE}      
	\begin{algorithmic}[1] 
		\footnotesize{
			\REQUIRE {Channel matrices $\mathbf{H}_{{\rm D},i}$, $\mathbf{H}_{{\rm C},i}$,  $\mathbf{H}_{{\rm S},i}$ $\forall i=1,2$, and iteration number $\mathcal{T}$.
			}

			\ENSURE {Hybrid beamforming $\mathbf{F}_i$ and $\mathbf{W}_i$, $\forall i=1,2$.} 
			
			\STATE{$\textbf{Initialize:}$  $\mathbf{F}_{{\rm RF},i}^{(0)}$ is randomly generated subject to the modulus-1 constraint, $\mathbf{F}_{{\rm BB},i}^{(0)}$ is randomly generated subject to $\left\Vert\mathbf{F}^{(0)}_{{\rm RF},i}\mathbf{F}^{(0)}_{{\rm BB},i}\right\Vert_F^2\leq \sigma_{p,i}^2$,
			iteration number $t=0$.
			}	
			\REPEAT		
			\STATE{Update $\mathbf{U}_j^{(t)}={\mathbf{H}}_{{\rm DC},i}\mathbf{F}_{{\rm RF},i}^{(t)}\mathbf{F}^{(t)}_{{\rm BB},i}\mathbf{F}_{{\rm BB},i}^{(t),H}\mathbf{F}_{{\rm RF},i}^{(t),H}{\mathbf{H}}_{{\rm DC},i}^H+{\mathbf{H}}_{{\rm S},j}\mathbf{F}_{{\rm RF},j}^{(t)}\mathbf{F}_{{\rm BB},j}^{(t)}\mathbf{F}_{{\rm BB},j}^{(t),H}\mathbf{F}_{{\rm RF},j}^{(t),H}{\mathbf{H}}_{{\rm S},j}^H+\sigma_{n,j}^2\mathbf{I}_{N_{j,r}}$.}
			\STATE{Obtain $\mathbf{W}_{{\rm BB},j}^{(t)}=\left(\mathbf{W}_{{\rm RF},j}^{(t),H}\mathbf{U}^{(t)}_j\mathbf{W}_{{\rm RF},j}^{(t)}\right)^{-1}\mathbf{W}_{{\rm RF},j}^{(t)}{\mathbf{H}}_{{\rm DC},i}\mathbf{F}^{(t)}_{{\rm RF},i}\mathbf{F}^{(t)}_{{\rm BB},i}$.
			}
		\STATE{Update $\mathbf{W}_{{\rm RF},j}^{(t)}$ according to proposition 2.
		}
		\STATE{Calculate $\mathbf{Q}_j^{(t)}=\mathbf{E}_j^{-1}$ according to Eqn. (\ref{EJ}).}
		
\STATE{Calculate $\widetilde{\mathbf{T}}^{(t)}_i=\mathbf{F}_{{\rm RF},i}^{(t),H}{\mathbf{H}}_{{\rm DC},i}^H\mathbf{W}_j^{(t)}
	\mathbf{Q}_j^{(t)}\mathbf{W}_j^{(t),H}{\mathbf{H}}_{{\rm DC},i}\mathbf{F}_{{\rm RF},i}^{(t)} + \mathbf{F}_{{\rm RF},i}^{(t),H}{\mathbf{H}}_{{\rm S},i}^H\mathbf{W}_i^{(t)}\mathbf{Q}_i^{(t)}\mathbf{W}_i^{(t),H}{\mathbf{H}}_{{\rm S},i}\mathbf{F}_{{\rm RF},i}^{(t)}$.}
\STATE{Update $\mathbf{F}_{{\rm BB},i}^{(t)}=\left(\widetilde{\mathbf{T}}_i^{(t)}+\widetilde{\mu}_i^{(t)}\mathbf{F}_{{\rm RF},i}^{(t),H}\mathbf{F}_{{\rm RF},i}^{(t)}\right)^{-1}\mathbf{F}_{{\rm RF},i}^{(t),H}{\mathbf{H}}_{{\rm DC},i}^H \mathbf{W}_j^{(t)}\mathbf{Q}_j^{(t)}$, where $\widetilde{\mu}_i^{(t)}$ is determined by the bisection algorithm to meet  $\left\Vert\mathbf{F}^{(t)}_{{\rm RF},i}\mathbf{F}^{(t)}_{{\rm BB},i}\right\Vert_F^2\leq \sigma_{p,i}^2$. }
	\STATE{Update $\mathbf{F}_{{\rm RF},i}^{(t)}$ according to proposition 2.
	}
		
		\STATE{$t=t+1$.}	
			\UNTIL{$t\geq\mathcal{T}$ is reached.}
		}
		
	\end{algorithmic}
\end{algorithm}

\vspace{-0.4cm}
\section{Simulation Results}\label{Sec6}

We assume that the two TX's pilot/transmit power is identical, so is the two RX's noise power, i.e., $\sigma_{p,1}^2=\sigma_{p,2}^2=\sigma_p^2$ and $\sigma_{n,1}^2=\sigma_{n,2}^2=\sigma_n^2$.
  The noise power is set as $-90$ dBm.
 All arrays are assumed to be uniform squared arrays, with
 $N_{i,t}=N_{j,r}=64, \forall i,j=1,2$, $L=256$, and $M_{i,t}=M_{j,r}=N_{st,i}=N_{st}$. The array spacing $D_0=20\lambda$ and relative angle $\omega=0$.
We assign the two TRXs the same training pilot length for channel estimation, i.e.,
  $N_{S,Y}=N_{S,X}=N_S$, $N_{D_j,Y}=N_{D_i,X}=N_D$, $N_{C_j,Y}=N_{C_i,X}=N_C$, $\forall i,j=1,2$. The look-ahead parameter $\mathcal{I}$ for LAOMP is set to $5$.
  For channel parameters, we employ the $28$ GHz mmWave channel setting in \cite{channel}. Following Table. I in \cite{channel}, we assume that path gains follow $\mathcal{CN}(0,\aleph10^{-0.1{\rm PL}})$, where $\aleph=\varepsilon_1^{1.8}10^{0.1\varepsilon_2}$ with $\varepsilon_1\sim \mathcal{U}(0,1)$ and $\varepsilon_2\sim \mathcal{CN}(0,16)$. Besides, ${\rm PL}=\zeta_1+10\zeta_2{\rm log}_{10}(d^\prime)+\mathcal{N}(0,\sigma) [\rm{dB}]$ with $d^\prime$ denoting the distance in meters, where $\zeta_1=72$, $\zeta_2=2.92$ and $\sigma=8.7$ for NLoS paths, and $\zeta_1=61.4$, $\zeta_2=2$ and $\sigma=5.8$ for LoS paths. All channels' number of paths is uniformly distributed in $[2,5]$. 
 Moreover, we fix the distance between the BS and the RIS as $45$ m, and assume that the distance in meter between the BS and the UE, and between the RIS and the UE are uniformly distributed in $[25,65]$ and $[1,20]$, respectively. Particularly, the one-way distance between the scatter and the BS/UE for SI NLoS channel model is uniformly distributed in $[15,30]$ meters.
  
\vspace{-0.4cm}
\subsection{Channel Estimation Error}

\begin{figure}
	\begin{minipage}[t]{0.496\linewidth}
		\centering
		\includegraphics[height=5.7cm,width=7.3cm]{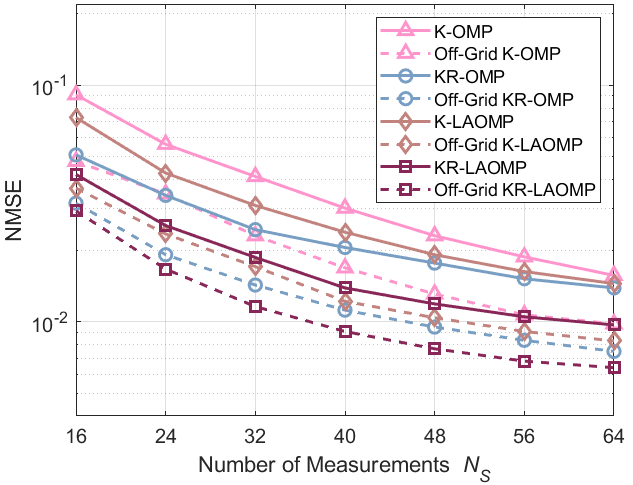}
		\caption{NMSE of estimated SI channel  versus $N_S$ with $\sigma_p^2=$ 30 dBm.}\label{CE-S}
	\end{minipage}%
	\begin{minipage}[t]{0.49\linewidth}
		\centering
		\includegraphics[height=5.7cm,width=7.3cm]{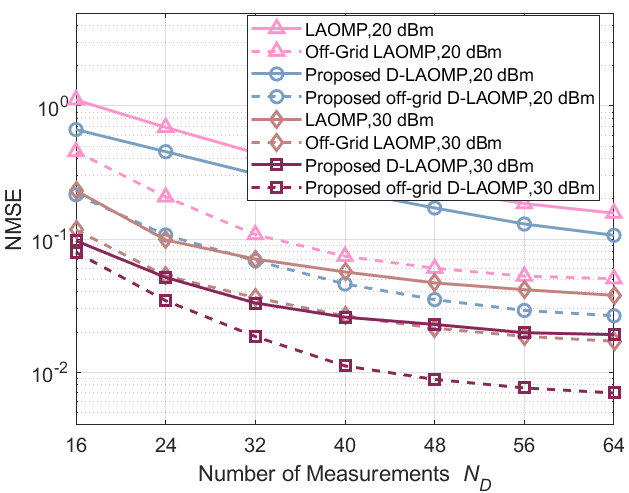}
		\caption{NMSE of estimated direct channel versus $N_D$ with $\sigma_p^2=$ 20 and 30 dBm.}\label{CE-D}
	\end{minipage}
	
\end{figure}
First, we use the normalized MSE (NMSE) to evaluate the channel estimation performance, which is defined as $\mathbb{E}\{\Vert\mathbf{H}-\widehat{\mathbf{H}}\Vert_F^2/\Vert\mathbf{H}\Vert_F^2\}$ with $\mathbf{H}$ and $\widehat{\mathbf{H}}$ denoting the true and estimated channels, respectively. 
In Fig. \ref{CE-S}, we exhibit the NMSE performance of the estimated UE's SI NLoS channel when $N_S$ ranging from 16 to 64, where $\sigma_p^2=30$ dBm. We denote OMP/LAOMP under the K-CS and KR-CS frameworks by K-OMP/K-LAOMP and KR-OMP/KR-LAOMP, respectively. It can be observed that KR-OMP/KR-LAOMP outperforms K-OMP/K-LAOMP since the former utilizes the angular-domain reciprocity to reduce the redundant dictionary size for better atom matching. In other words, K-CS is easier to fall into the local optimal solution than KR-CS. Besides, KR-CS is faster than K-CS, which is shown in later complexity analysis. Moreover, we can see that LAOMP-based methods are better than OMP-based methods in terms of NMSE due to the look-ahead procedure, and off-grid methods achieve a significant performance improvement by addressing the basis mismatch problem.
 In later simulations, OMP is not applied for the benchmark.

Next, we evaluate our proposed D-LAOMP for direct channel estimation. To intuitively reflect our proposed method's advantage, we assume the SI channel is perfect such that we can completely eliminate the influence of the SI pilot when estimating the direct channel. Fig. \ref{CE-D} shows the NMSE performance of the two methods with changing pilot length ($N_D\in[16,64]$), and $\sigma_p^2=$ 20 and 30 dBm, where LAOMP/off-grid LAOMP follows the normal channel estimation procedure, i.e., estimating the DL or UL channel with LAOMP/off-grid LAOMP algorithms, respectively. It is shown that our proposed D-LAOMP and off-grid D-LAOMP methods for joint DL-UL channel estimation have a good performance improvement. From another perspective, when a certain NMSE is specified, our proposed frameworks can conduct channel estimation with less pilot overhead. 
\begin{figure}
	\begin{minipage}[t]{0.496\linewidth}
		\centering
		\includegraphics[height=5.6cm,width=7.3cm]{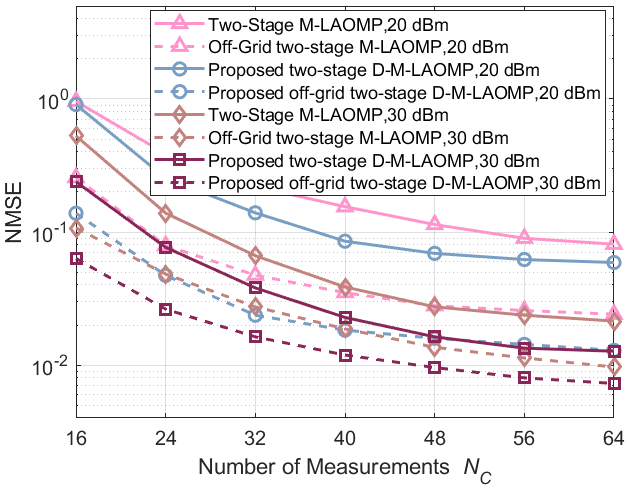}
		\caption{NMSE of estimated BS's and UE's AoA/AoD versus $N_C$ .}\label{CE-C}
	\end{minipage}%
	\begin{minipage}[t]{0.49\linewidth}
		\centering
		\includegraphics[height=5.6cm,width=7.3cm]{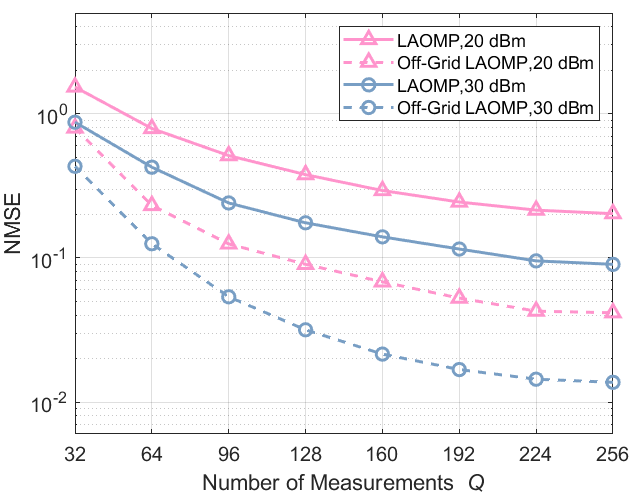}
		\caption{NMSE of estimated RIS's AoA/AoD versus $Q$.}\label{CE-AC}
	\end{minipage}
	
\end{figure}

In Fig. \ref{CE-C}, the NMSE performance of our proposed two-stage D-M-LAOMP method for BS's and UE's parameter estimation is shown, where ${N}_C$ ranges from 16 to 64, and pilot power is set to 20 and 30 dBm, respectively. Since the BS's and UE's parameters are estimated with the fixed RIS phase matrix $\mathbf{V}_0$, here the NMSE is expressed as $\mathbb{E}\{\Vert\mathbf{H}_{\rm R}\mathbf{V}_0\mathbf{H}_{\rm T}-\widehat{\mathbf{H}_{\rm R}\mathbf{V}_0\mathbf{H}_{\rm T}}\Vert_F^2/\Vert \mathbf{H}_{\rm R}\mathbf{V}_0\mathbf{H}_{\rm T}\Vert_F^2\}$. From Fig. \ref{CE-C}, it can be shown that our proposed two-stage D-M-LAOMP/off-grid D-M-LAOMP has a significant performance improvement compared with two-stage M-LAOMP/off-grid M-LAOMP. This further demonstrates the advantage of joint DL-UL training. Moreover, the NMSE performance of RIS's AoA/AoD estimation (angular-domain cascaded channel estimation) is shown in Fig. \ref{CE-AC}, where the NMSE of the angular-domain cascaded channel is defined as $\mathbb{E}\{\Vert\bm{\Xi}-\widehat{\bm{\Xi}}\Vert_F^2/\Vert\bm{\Xi}\Vert_F^2\}$. It can be observed that our proposed off-grid LAOMP can achieve a good NMSE performance with an appropriate number of measurements $Q$.
\vspace{-0.4cm}
\subsection{Spectral Efficiency}
\begin{figure}
	\begin{minipage}[t]{0.496\linewidth}
		\centering
		\includegraphics[height=5.6cm,width=7.3cm]{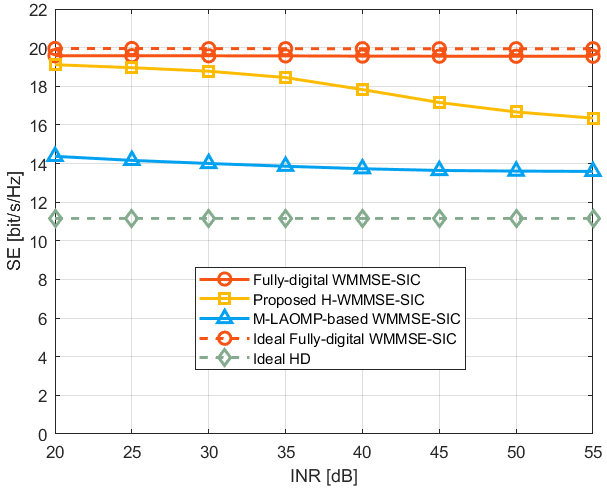}
		\caption{SE versus INR with $\sigma_p^2=10$ dBm.}\label{SE-INR10}
	\end{minipage}%
	\begin{minipage}[t]{0.49\linewidth}
		\centering
		\includegraphics[height=5.6cm,width=7.3cm]{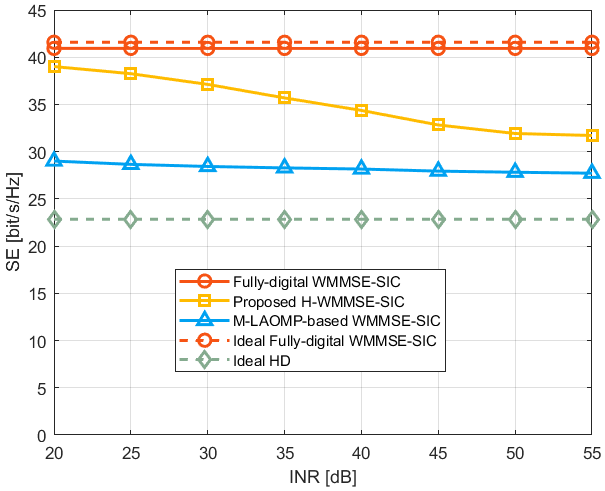}
		\caption{SE versus INR with $\sigma_p^2=20$ dBm.}\label{SE-INR20}
	\end{minipage}
	
\end{figure}

To show the effectiveness of our proposed H-WMMSE-SIC method with perfect channels, the following benchmarks are employed: fully-digital WMMSE-SIC, M-LAOMP-based WMMSE-SIC, proposed H-WMMSE-SIC, ideal FD, and ideal HD. For concreteness, M-LAOMP-based WMMSE-SIC aims to directly decouple the fully-digital WMMSE precoder/combiner with M-LAOMP, this decoupling way is similar to sparse hybrid beamforming \cite{HB1}. Ideal FD means there is no SI between the TRX arrays, i.e., the SI channel is set to a null matrix. Ideal HD denotes the SE of UL or DL. Considering the energy fairness mechanism, FD and HD have the same energy, such that the power consumption of HD is set to $2\sigma_p^2$. 
 As our previous statement, passive SIC techniques and the analog/digital circuit can be helpful for SIC. Thus, we consider a changing residual SI to evaluate the impact of different SI power on our proposed method. 
 Recalling the SI LoS channel's path loss is $\gamma^{\rm LoS}$, we define the interference-to-noise ratio (INR) as $(\gamma^{\rm LoS})^2/\sigma_n^2$. Here we just adjust the SI LoS channel's path loss according to the INR, the SI NLoS channel's path loss follows the channel parameter setting described before.
 
 In Fig. \ref{SE-INR10} and \ref{SE-INR20}, we plot the SE performance with INR ranging from 20 to 55 dB, where the transmit power $\sigma_p^2$ is fixed as 10 and 20 dBm, and the number of data streams $N_{st}$ is set to $4$. 
 It is shown that fully-digital WMMSE-SIC can approach its ideal case without the SI, demonstrating the excellence of WMMSE used for SIC. A direct solution for hybrid beamforming, i.e., decoupling the fully-digital WMMSE-SIC solution into the digital and analog beamforming with M-LAOMP, has much performance loss since 1) it doesn't consider the optimal solution for analog beamforming and digital beamforming separately, 2) limited number of RF chains will significantly impact the performance, and 
  3) the two decoupling errors of fully-digital precoder $\mathbf{F}$ and fully-digital combiner $\mathbf{W}$ will affect the performance together. 
  In constrast, our proposed H-WMMSE-SIC, which iteratively updates analog/digital beamforming in the WMMSE procedure for an optimal solution in each iteration, outperforms M-LAOMP-based WMMSE-SIC and approaches fully-digital WMMSE at an appropriate INR. Moreover, the SE of ideal HD is about half of that of ideal fully-digital WMMSE-SIC, and it can exceed half owing to the energy fairness mechanism.
 
\begin{figure}
	\begin{minipage}[t]{0.496\linewidth}
		\centering
		\includegraphics[height=5.64cm,width=7.3cm]{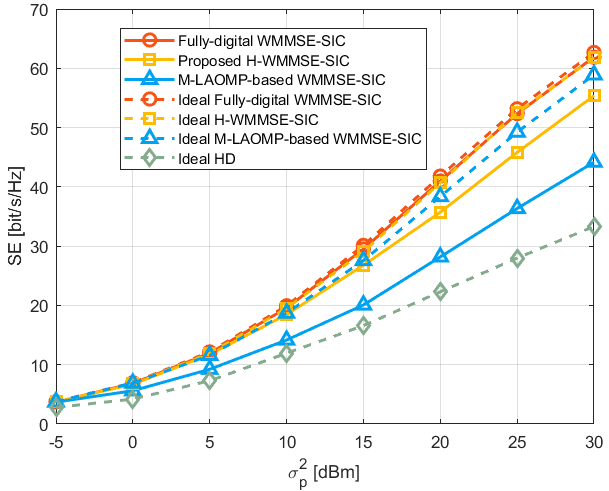}
		\caption{SE versus $\sigma_p^2$ with INR $=35$ dB and $N_{st}=4$. }\label{SE-dbm}
	\end{minipage}%
	\begin{minipage}[t]{0.49\linewidth}
		\centering
		\includegraphics[height=5.6cm,width=7.3cm]{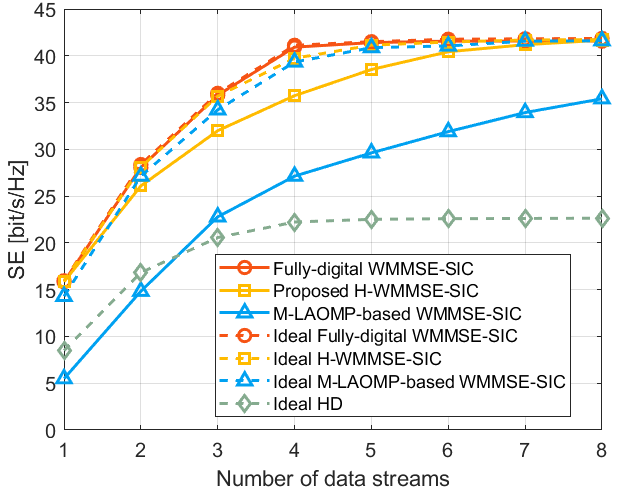}
		\caption{SE versus $N_{st}$ with INR $=35$ dB and $\sigma_p^2=20$ dBm.}\label{SE-stream}
	\end{minipage}
	
\end{figure}
 Fig. \ref{SE-dbm} exhibits the SE performance with $\sigma_p^2$ varying from $-5$ to $30$ dBm, where INR and $N_{st}$ are set as 35 dB and $4$, respectively. As the power $\sigma_p^2$ increases, the SE difference between ideal FD and ideal HD becomes large, as well as between our proposed H-WMMSE-SIC and M-LAOMP-based WMMSE-SIC. On the other hand, we can see these FD methods under ideal conditions have a good performance, and the gap among them is small. 
  To explore the impact of the number of transmit/receive data streams $N_{st}$ on different schemes, we plot in Fig. \ref{SE-stream} the SE curves of different schemes with the changing $N_{st}$, where $\sigma_p^2$ and INR are set to 20 dBm and 35 dB, respectively. It can be observed that as $N_{st}$ becomes larger, the gap between the fully-digital beamforming and hybrid beamforming becomes smaller. This is because the restricted available spatial rank limits the spatial capacity, large $N_{st}$ has little impact on the growth of SE. 
  In contrast, hybrid beamforming will approach fully-digital beamforming
  due to the assumption that the number of data streams equals to that of RF chains. From Fig. \ref{SE-stream}, our proposed H-WMMSE-SIC can achieve a near-optimal SE compared to fully-digital WMMSE-SIC when the number of RF chains is larger than $5$. However, the direct solution, M-LAOMP-based WMMSE-SIC, requires a large number of RF chains to achieve a near-optimal solution.
 
 \vspace{-0.4cm}
 \subsection{Computational Complexity Analysis}
For one MMV sparse recovery form $\mathbf{Y}=\mathbf{AX}$ with $\mathbf{A}\in\mathbb{C}^{M\times N}$ and $\mathbf{X}\in\mathbb{C}^{N\times R}$, the per-iteration complexity (PIC) of OMP and LAOMP with $\mathcal{I}$ look-ahead loops are $\mathcal{O}(MN)$ and $\mathcal{O}(\mathcal{I}MN)$, respectively. Since the PIC difference between OMP and LAOMP is only the constant $\mathcal{I}$, we will give the PIC of our proposed channel estimation schemes with OMP for clarity.
 Considering Eqn. (\ref{S2_CS}) with KR-CS for SI channel estimation, the PIC is $\mathcal{O}(N_{S,X}N_{S,Y}G)$. However, using K-CS for Eqn. (\ref{YS2}), the PIC will be $\mathcal{O}(N_{S,X}N_{S,Y}G^2)$. This is because the angular-domain reciprocity helps SI channel estimation remove the redundant information. In other words, KR-CS is the special case of K-CS when determining that non-zero elements are on the diagonal of the matrix to be recovered. With Eqn. (\ref{D_CS}), the PIC of direct channel estimation can be easily obtained as $\mathcal{O}({\rm max}\{N_{D_2,X}N_{D_1,Y}G^2,N_{D_1,X}N_{D_2,Y}G^2\})$. For cascaded channel estimation, the first stage of BS's and UE's channel parameter estimation, i.e., solving Eqn. (\ref{dmcs1}), dominates the PIC complexity of $\mathcal{O}({\rm max}\{N_{C_1,Y}N_{C_2,X}G,N_{C_1,X}N_{C_2,Y}G\})$. Furthermore, the complexity of off-grid methods after CS recovery can refer to \cite{OG1,OG2}. 
 
 For beamforming, passive beamforming only needs to be executed once with the inverse operation of $\mathcal{O}(L^3)$. Then we analyze the complexity of hybrid beamforming with PIC, here PIC refers to a loop of Algorithm \ref{HWMMSE}. Suppose that $N_{i,t}=N_{j,r}=\mathring{N}$ and $M_{i,t}=M_{i,r}=\mathring{M}$ for $\forall i,j=1,2$, and $\mathring{M}\ll\mathring{N}$. Looking at Algorithm \ref{HWMMSE}, \emph{Step 3}, \emph{4}, \emph{6}, and \emph{7} incur a PIC of $\mathcal{O}(\mathring{N}^2\mathring{M})$ by computing the product of one $(\mathring{N}\times\mathring{N})$-dimensional channel matrix and one $(\mathring{N}\times\mathring{M})$-dimensional matrix. Particularly, according to proposition 2, \emph{Step 5} and \emph{9} result in a PIC of $\mathcal{O}(\mathcal{X}\mathring{M}^2\mathring{N}^2)$
  due to the computation of $\nu_{k_1,k_2}$ in each iteration. Finally, \emph{Step 8} requires a PIC of $\mathcal{O}(\mathcal{V}\mathring{N}^2\mathring{M})$ with $\mathcal{V}$ denoting the number of performing the bisection algorithm to meet the power constraint. Overall, the whole PIC of Algorithm \ref{HWMMSE} is $\mathcal{O}({\rm max}\{\mathcal{X}\mathring{M}^2\mathring{N}^2,\mathcal{V}\mathring{N}^2\mathring{M}\})$.
 \vspace{-0.4cm}
\section{Conclusions}\label{Sec7}
The RIS-aided mmWave FD technique has a great potential in achieving high SE and low latency. In this regard, this paper has proposed efficient methods regarding channel estimation, passive beamforming, and hybrid beamforming. Unlike the usage of the traditional channel matrix reciprocity in some papers, we have demonstrated the existence of the angular-domain reciprocity inherent in mmWave FD systems, which is more appropriate for building spatial correlations between the TX and RX arrays. Based on this, SI channel estimation with KR-CS is developed for faster running time than that with K-CS. 
 Additionally, D-CS and two-stage D-M-CS frameworks are proposed for direct and cascaded channel estimation, respectively. Under the proposed CS frameworks, different variant LAOMP algorithms and their off-grid solutions are developed. 
  Simulations have exhibited that they outperform the conventional estimation methods without joint signal processing (i.e., the angular-domain reciprocity is not considered). Moreover, the optimization of passive and hybrid beamforming has been investigated by maximizing the SE while suppressing the SI. First, using the angular-domain cascaded channel, passive beamforming is optimized for improving the cascaded channel capacity. Along with this, a H-WMMSE-SIC algorithm, mining  the hybrid beamforming design and the interference elimination advantage of WMMSE, is proposed for maximizing the spatial multiplex gain while minimizing the SI. Simulations have shown its excellence in terms of SIC with a high INR. Meanwhile, we show that H-WMMSE-SIC has a low computational complexity.

\begin{appendices}
	\vspace{-0.4cm}
\section{  }\label{appendixA}
Recalling Eqn. (\ref{Rr}), we have
\begin{equation}
	\begin{aligned}
R_{r,(n_r^z,n_r^y)}=&\sqrt{R^2-2R\left(n_r^yd\Psi^{\rm azi}+((D_0+N_t^zd)+n_r^zd)\Psi^{\rm ele} \right)+(n^y_rd)^2+((D_0+N_t^zd)+n_r^zd)^2}\\
\approx&  R-n_r^yd\Psi^{\rm azi}-((D_0+N_t^zd)+n_r^zd)\Psi^{\rm ele}+\frac{(n_r^yd)^2}{2R}\left(1-(\Psi^{\rm azi})^2\right) \\ & +\frac{((D_0+N_t^zd)+n_r^zd)^2}{2R}(1-\left(\Psi^{\rm ele})^2\right)+\frac{n_r^yd((D_0+N_t^zd)+n_r^zd)\Psi^{\rm ele}\Psi^{\rm azi}}{R} + ... \ .
	\end{aligned}
\end{equation}

Ignoring items above the square, the phase error is less than or equal to
$
\frac{2\pi}{\lambda}(\frac{(n_r^yd)^2}{2R}+\frac{((D_0+N_t^zd)-n_r^zd)^2}{2R})
$. Thus, when
\begin{equation}
\begin{aligned}
\frac{(n_r^yd)^2}{2R}+\frac{((D_0+N_t^zd)+n_r^zd)^2}{2R} & \leq \frac{(n_r^yd)^2}{2R_{\rm min}}+\frac{((D_0+N_t^zd)+n_r^zd)^2}{2R_{\rm min}} \\ &    \leq \frac{(N_r^yd)^2}{2R_{\rm min}}+\frac{((D_0+N_t^zd)+N_r^zd)^2}{2R_{\rm min}} \\ &\leq \frac{\lambda}{2\pi}\delta_{\rm max},
\end{aligned}
\end{equation}
we obtain $D_0\leq \sqrt{\frac{\lambda}{\pi}R_{\rm min}\delta_{\rm max}-(N_r^yd)^2}-N_t^zd-N_r^zd$. The proof is complete.
\vspace{-0.4cm}
\section{  }\label{appendixB}
Here we give the proof of proposition 2.
First, we define the objective function $f(\mathbf{R})\triangleq{\rm Tr}\left(
\mathbf{B}^H\mathbf{R}^H\mathbf{U}\mathbf{R}\mathbf{B}\right)+2\Re\left\{{\rm Tr}\left\{\mathbf{B}^H\mathbf{R}^H\mathbf{F}\right\}\right\}$, which can be expressed as a quadratic form w.r.t. $[\mathbf{R}]_{k_1,k_2}$, i.e., $f([\mathbf{R}]_{k_1,k_2})\triangleq\kappa\vert[\mathbf{R}]_{k_1,k_2}\vert^2-2\Re\{\nu_{k_1,k_2}^*[\mathbf{R}]_{k_1,k_2}\}$, where $\kappa_{k_1,k_2}$ and $\nu_{k_1,k_2}$ are derived as follows. According to \cite{matrix}, we can obtain
\begin{equation}
\frac{\partial f(\mathbf{R})}{\mathbf{R}^*}=\mathbf{URB}\mathbf{B}^H-\mathbf{F}\mathbf{B}^H,
\ \
\frac{\partial f([\mathbf{R}]_{k_1,k_2})}{[\mathbf{R}^*]_{k_1,k_2}}=\kappa_{k_1,k_2}[\mathbf{R}]_{k_1,k_2}-\nu_{k_1,k_2}.
\end{equation}
Due to $[\frac{\partial f(\mathbf{R})}{\mathbf{R}^*}]_{k_1,k_2}=\frac{\partial f([\mathbf{R}]_{k_1,k_2})}{[\mathbf{R}^*]_{k_1,k_2}}$, it can be known that 
\begin{equation}
	\begin{aligned}
[\mathbf{URB}\mathbf{B}^H-&\mathbf{F}\mathbf{B}^H]_{k_1,k_2}=\sum_{k_3}\sum_{k_4}[\mathbf{U}]_{k_1,k_3}[\mathbf{R}]_{k_3,k_4}[\mathbf{BB}^H]_{k_4,k_2}-[\mathbf{FB}^H]_{k_1,k_2}\\
&=[\mathbf{U}]_{k_1,k_1}[\mathbf{R}]_{k_1,k_2}[\mathbf{BB}^H]_{k_2,k_2}+\sum_{k_3\neq k1}\sum_{k_4\neq k2}[\mathbf{U}]_{k_1,k_3}[\mathbf{R}]_{k_3,k_4}[\mathbf{BB}^H]_{k_4,k_2}-[\mathbf{FB}^H]_{k_1,k_2}\\
&=\kappa_{k_1,k_2}[\mathbf{R}]_{k_1,k_2}-\nu_{k_1,k_2}.
	\end{aligned}
\end{equation}
Hence we have $\kappa_{k_1,k_2}=[\mathbf{U}]_{k_1,k_1}[\mathbf{BB}^H]_{k_2,k_2}$, and $\nu_{k_1,k_2}=\sum_{k_3\neq k1}\sum_{k_4\neq k2}[\mathbf{U}]_{k_1,k_3}[\mathbf{R}]_{k_3,k_4}[\mathbf{BB}^H]_{k_4,k_2}-[\mathbf{FB}^H]_{k_1,k_2}$. Recalling the optimization problem ${\rm min} f(\mathbf{R})$, subject to $\vert[\mathbf{R}]_{k_1,k_2}\vert=1$, $\forall k_1,k_2$, which is equivalent to 
\begin{equation}
	\begin{aligned}
	\underset{[\mathbf{R}]_{k_1,k_2}}{\rm arg \ min} \ -2\Re\{\nu_{k_1,k_2}^*[\mathbf{R}]_{k_1,k_2}\} , \ {\rm s.t.} \	\vert[\mathbf{R}]_{k_1,k_2}\vert=1.
\end{aligned}
\end{equation}
\end{appendices}
We can get the optimal solution $[\mathbf{R}]_{k_1,k_2}=\frac{\nu_{k_1,k_2}}{\vert\nu_{k_1,k_2}\vert}$. With the manner that iteratively updates each element while fixing other elements, all elements of $\mathbf{R}$ can be optimized. Overall, $\mathcal{X}$ loops are performed, with each loop optimizing all elements of $\mathbf{R}$, for performance guarantee.

\bibliographystyle{IEEEtran}
\bibliography{reference.bib}

\vspace{12pt}

\end{document}